\documentclass[titlepage,jcp,aps,preprint,groupedaddress,nofootinbib,nobibnotes,superscriptaddress]{revtex4-1}

\usepackage[dvipdfmx]{color}
\usepackage[hiresbb]{graphicx}
\usepackage{amssymb,amsfonts,amsmath}
\usepackage{amsfonts}
\usepackage{newtxtext,newtxmath}
\usepackage{amsmath,amsbsy,amstext}
\usepackage{mathrsfs}
\usepackage{array}
\usepackage{multirow}
\usepackage{tabularx}
\usepackage{booktabs}
\usepackage{subcaption} 
\usepackage{mathtools}   
\usepackage{subcaption,dcolumn}
\newcolumntype{d}[1]{D..{#1}}
\usepackage{array}
\newcolumntype{P}[1]{>{\centering\arraybackslash}p{#1}}

\begin{document}

\begin{abstract}
In the molecular dynamics calculations for the free energy of ions and ionic molecules, we often encounter wet charged molecular systems where electrical neutrality condition is broken. This causes a problem in the evaluation of electrostatic interaction under periodic boundary condition. A standard remedy for the problem is to consider a hypothetical homogeneous background charge density to neutralize the total system. Here, we present a new expression for the evaluation of electrostatic interactions for the system including the background charge by fast multipole method (FMM). Further, an efficient scheme to evaluate solute-solvent interaction energy by FMM has been developed to reduce the computation of far-field part. 
We have calculated hydration free energy of ions,  Mg\(^{2+}\), Na\(^{+}\), and Cl\(^{-}\) dissolved in 
neutral solvent using the new expression. 
The calculated free energy showed a good agreement with the result using  well-established particle mesh Ewald method, demonstrating the validity of the present expression in the framework of FMM.
An advantage of the present scheme is in an efficient free energy calculation of a large-scale charged systems (particularly over million particles) based on highly parallel computations. 
\end{abstract}

\title{Exact electrostatic energy calculation for charged systems neutralized by uniformly distributed background charge using fast multipole method and its application to efficient free energy calculation}

\newcommand{\footremember}[2]{%
    \fo

otnote{#2}
    \newcounter{#1}
    \setcounter{#1}{\value{footnote}}%
}
\newcommand{\footrecall}[1]{%
    \footnotemark[\value{#1}]%
} 

\author{Ryo Urano}
\affiliation{Department of Materials Chemistry, Nagoya University, Nagoya 464-8603, Japan}

\author{Wataru Shinoda}
\affiliation{Department of Materials Chemistry, Nagoya University, Nagoya 464-8603, Japan}
\affiliation{Center for Computational Science, Graduate School of Engineering, Nagoya University, Nagoya 464-8603, Japan}  %

\author{Noriyuki Yoshii}
\altaffiliation{Corresponding author}
\altaffiliation{Phone/Fax: +81-52-788-6213, E-mail: yoshii@ccs.engg.nagoya-u.ac.jp (N. Y. )}
\affiliation{Department of Materials Chemistry, Nagoya University, Nagoya 464-8603, Japan}
\affiliation{Center for Computational Science, Graduate School of Engineering, Nagoya University, Nagoya 464-8603, Japan}  %

\author{Susumu Okazaki}
\altaffiliation{Corresponding author}
\altaffiliation{Phone/Fax: +81-52-789-5829, E-mail: okazaki@chembio.nagoya-u.ac.jp (S. O. )}
\affiliation{Department of Materials Chemistry, Nagoya University, Nagoya 464-8603, Japan}
\affiliation{Center for Computational Science, Graduate School of Engineering, Nagoya University, Nagoya 464-8603, Japan}  %


\maketitle

\section*{Introduction}

Molecular dynamics (MD) calculation is a powerful tool to investigate thermodynamic properties, structure, and dynamics of molecular assemblies at an atomistic level. A large number of MD calculations have been widely performed for various systems such as pure water, electrolyte solution, and biomolecular solution where protein molecules and lipid membranes are dissolved. In these systems, intermolecular interaction of water, ions, and other solute molecules are expressed by  electrostatic interaction as well as Lennard-Jones interaction \cite{allen2017computer,tuckerman2010statistical,frenkel2001understanding}.

The electrostatic interaction is a long-ranged one and occupies a large portion of MD calculation time. In a large-scale system containing millions of atoms or more, the calculation time for the electrostatic interaction calculation is dominant in the MD calculation time. In order to evaluate the electrostatic interaction efficiently, the particle mesh Ewald (PME) method \cite{essmann1995smooth} was developed and is adopted by many MD programs \cite{Berendsen1995,lindahl2001gromacs,van2005gromacs,hess2008gromacs,pronk2013gromacs,abraham2015gromacs,brooks1983charmm,brooks2009charmm,jung2015genesis,kobayashi2017genesis}. In the PME method, arithmetic operations increase with \(O\)( \(N\) log \(N\)) with increasing number of atoms \(N\).  However, the PME method requires execution of  Fast Fourier Transform (FFT) frequently, in which all-to-all communication wastes huge communication time in a highly parallel environment \cite{phillips2014mapping}. Thus, the cost of PME is very high for the highly parallel computation. In order to avoid this, a new algorithm realizing high parallelization efficiency with low communication load is important.

So far, algorithms such as Wolf's method \cite{wolf1999exact}, zero multipole method \cite{fukuda2013zero,fukuda2014zero}, and multilevel summation method \cite{hardy2015multilevel} have been developed, which can perform electrostatic interaction with \(O(N)\). However, it should be noted that these non-Ewald methods suffer from truncation of long-ranged contribution from distant image cells.
Fast multipole method (FMM) \cite{greengard_new_1997,greengard1988rapid} is a rigorous one where the interactions strictly summed up to the infinite image cells.

The FMM was first developed by Greengard \(et~al\). for an isolated systems and was immediately extended to systems in the periodic boundary condition \cite{figueirido_large_1997,amisaki_precise_2000,schmidt1991implementing}, combined with RESPA algorithm \cite{zhou1995new}. Since the FMM is a somewhat complicated algorithm, it has not been widely used yet in MD software. However, a simpler expression by solid harmonics than that by original spherical harmonics is now given \cite{epton1994ma,van1998shift,nitadori_particle_2014}. The FMM is numerically rigorous based on an analytic formula such that the calculation accuracy can be controlled by the degree of expansion of the series of the harmonics 
while that of the Ewald method is similarly controlled by the cutoff distance in the real space together with the order of FFT calculation of in the reciprocal space. 
In our previous study, we showed good conservation of hamiltonian during the large-scale MD calculation for 10 million atom systems with the expansion degree of four \cite{andoh2013modylas}. 

The FMM has been implemented in our highly parallelized general-purpose MD calculation software MODYLAS \cite{andoh2013modylas}. In the FMM program, since the regional decomposition technique of the system is used for the parallelization, communications are required just between adjacent nodes, suitable to highly parallel computations. Thus, since the method is free from all-to-all communications, time required for the communications in the FMM is significantly short compared with the PME in a highly parallel environment.

However, there are few applications of FMM to MD calculation for the condensed matter systems. Especially, no free energy calculations have been reported so far using FMM. Free energy is quite crucial when we evaluate the stability and reactivity of the equilibrium systems \cite{atkins2011physical}. Many of the free energy calculation methods have been developed so far based on MD calculation\cite{frenkel2001understanding}. In particular, particle insertion method, the thermodynamic integration method (TI), and free energy perturbation method are well known as high-accuracy calculation methods. In these methods, the free energy difference between a reference system and a target system has been evaluated by the difference in interaction energy itself between the two systems or by the first derivative of interaction function with respect to an adopted parameter. 

For a net charged solute such as ion and  protein, an additional care must be taken for the free energy calculation. 
For example, in the hydration process of ions from vacuum to aqueous solution in the thermodynamic integration method, the electrically neutral condition fails for either of the vacuum and the aqueous solution. To avoid the divergence of the calculated potential energy in the Ewald methods, electrical neutrality is kept by adding a uniform background charge density with opposite sign \cite{hummer1996free,hub2014quantifying}.  However, no analytical expression by FMM has not yet been given for the energies coming from coulombic interaction of the background charge density. The expression is very important when we use the FMM for the free energy calculation where molecules of interest have a charged state.

In this study, we first apply the background charge method to the free energy calculation based on FMM presenting a new FMM expression for the solute-solvent interaction energy of the system. 
The results were compared with those obtained by conventional PME method to discuss its accuracy.

\section*{Theory}

\subsection*{Electrostatic interaction calculation by FMM for systems with background charge}

In general, point charges (PC) in a unit cell satisfy the electrically neutral condition. In this case,  electrostatic interactions in the unit cell under periodic boundary condition can be evaluated by conventional FMM \cite{greengard_new_1997,amisaki_precise_2000}. The interaction of a tagged particle in a divided cell of interest (focused cell) is evaluated by the sum of the interactions with particles in the neighboring cells (near field: NF) and  in the far distant cells (far field: FF) as shown in Fig. \ref{Fig1}. Interaction of the tagged particle with particles in the NF is evaluated directly by coulombic pair interactions. Interaction with the particles in the FF is evaluated by the lattice sum of the multipole moment of the image cells as given in Eq. (\ref{eq:12}) shown below. This lattice sum can be divided into a real space term (\ref{eq:12a}), reciprocal lattice space term (\ref{eq:12b}), and self  term (\ref{eq:12c}). The reciprocal space term is composed of two terms with wave vector \(\boldsymbol{k} \ne \boldsymbol{0}\) and \(\boldsymbol{k} = \boldsymbol{0}\). The \(\boldsymbol{k} = \boldsymbol{0}\) term and self term disappear when the electrical neutrality is satisfied.

When the system does not satisfy the electrical neutrality, the FMM calculation suffers from divergence of the \(\boldsymbol{k} = \boldsymbol{0}\) term. In order to avoid this, a uniform background charge (BC) density is introduced to satisfy the electrical neutrality of the system.
Then, the electrostatic interactions consist of PC-PC, PC-BC, and BC-BC interactions. The formula for each interaction will be presented below and is listed in Table \ref{Table2}.
The PC-PC interaction in the FF diverges when the neutral condition is not satisfied.
Further, sum of the PC-BC and BC-BC interactions analytically obtained using the Ewald method (See Eq. (\ref{eq:surfcharged3}) and Appendix  \ref{ewaldcharge}). 
The sum of PC-BC and BC-BC interactions also diverges alone. However, these two divergent terms cancel out each other, and, as a result, a new quadrupole surface term remains as shown later.

In the following sections, contribution from the image  FF  cells is first presented in Eq. (\ref{eq:12}) with  divergent terms. We then show that the self term should be added to the equation under the non-electrically neutral condition, which disappears under the electrically neutral condition.  We also show that at \(\boldsymbol{k}=\boldsymbol{0}\), the PC-PC, PC-BC, and BC-BC interactions result in a new quadrupole surface term. Thus, we must take account of several new terms newly appear in the calculation of FMM with BC neutralization.

\subsection*{Contribution from the far-field image cells}

FMM for net charged systems consisting of particle PC ($Q_{\mathrm{C}}\neq 0$, where $Q_{\mathrm{C}}$ is the sum of the PC) and hypothetical BC is outlined. For the introduction to the conventional FMM \cite{greengard1988rapid}, see Appendix \ref{APPDConvFMM}.
In the FMM under periodic boundary condition, 
irregular solid harmonics $\boldsymbol{\mathcal{I}}_{\lambda }^{\eta }$ with degree $\lambda$ and order $\eta$ is summed over the image cells.
The summation $\sum _{\boldsymbol{\nu }\neq \boldsymbol{0}}\boldsymbol{\mathcal{I}}_{\lambda }^{\eta }\left(\boldsymbol{{\boldsymbol { r}}}_{\boldsymbol{\nu }}\right)$ for the net charged systems 
is evaluated using incomplete gamma function, $\gamma (\alpha ,x)=\int _{0}^{x}e^{-t}t^{\alpha -1}\mathrm{d}t$, 
and its complementary function, $\Gamma (\alpha ,x)=\int _{x}^{\infty }e^{-t}t^{\alpha -1}\mathrm{d}t$, to split it into real-space term and reciprocal-space term 
like Ewald summation in Eqs. (\ref{eq:Ereal}) and (\ref{eq:Eself}) in Appendix \ref{ewaldcharge} \cite{schmidt1991implementing,amisaki_precise_2000,zhou1995new}. Then, $\sum _{\mathbf{\pmb{\nu} }\neq \mathbf{0}}\boldsymbol{\mathcal{I}}_{\lambda }^{\eta }\left(\boldsymbol{r}_{\mathbf{\pmb{\nu} }}\right)$ can be written by
\begin{subequations}
\label{eq:12}
\begin{eqnarray}
\sum _{\mathbf{\pmb{\nu} }\neq \mathbf{0}}\boldsymbol{\mathcal{I}}_{\lambda }^{\eta }\left(\boldsymbol{r}_{\mathbf{\pmb{\nu} }}\right)&=&\sum _{\mathbf{\pmb{\nu} }\neq \mathbf{0}}^{\left| \mathbf{r}_{\mathbf{\pmb{\nu} }}\right| \leq r_{\nu \max }}\boldsymbol{\mathcal{I}}_{\lambda }^{\eta }\left(\boldsymbol{r}_{\mathbf{\pmb{\nu} }}\right)\frac{\Gamma \left(\lambda +\frac{1}{2},\kappa ^{2}{{\boldsymbol {r}}_{\mathbf{\pmb{\nu} }}}^{2}\right)}{\Gamma \left(\lambda +\frac{1}{2}\right)}  \label{eq:12a}  \\
&+&\frac{\pi ^{\frac{3}{2}}i^{\lambda }}{2^{\lambda - 2}\Gamma \left(\lambda +\frac{1}{2}\right)V}\sum _{\mathbf{k } = \mathbf{0}}^{\left| \boldsymbol{k}_{\mathbf{}}\right| \leq k_{\max }}\boldsymbol{\mathcal{I}}_{\lambda }^{\eta }\left(\boldsymbol{k}_{\mathbf{}}\right)k_{\mathbf{ }}^{2\lambda - 1}e^{- \frac{{k_{\mathbf{ }}}^{2}}{4\kappa ^{2}}}  \label{eq:12b}  \\
&-& \frac{2\kappa }{\sqrt{\pi }}\delta _{\lambda 0}\delta _{\eta 0},   \label{eq:12c}
\end{eqnarray}
\end{subequations}
where $\boldsymbol{{\boldsymbol {r}}}_{\boldsymbol{\nu }}=\boldsymbol{\nu }b$ is the centered position of the $\boldsymbol{\nu }$-th image cell and $b$ is the side length of the unit cell, $\Gamma (\alpha )=\int _{0}^{\infty }e^{-t}t^{\alpha -1}\mathrm{d}t$ is the gamma function, and $\kappa $ is the splitting parameter. 
The summation for $\boldsymbol{\nu }$ and $\boldsymbol{k}$ in terms (\ref{eq:12a}) and (\ref{eq:12b}) is taken over the range of $\left| \boldsymbol{r}_{\mathbf{\pmb{\nu} }}\right| \leq r_{\nu \max }$ and $\left| \boldsymbol{k}\right| \leq k_{\max }$, respectively, where $r_{\nu \max }$ and $k_{\max }$ are parameters which determine calculation accuracy. The first (\ref{eq:12a}) and the second (\ref{eq:12b}) terms are real space and reciprocal space terms, respectively. The third (\ref{eq:12c}) term is the self term and can be derived in a similar way to the Ewald method. The derivation of the self term is shown in Appendix \ref{selfterm}.  

\subsection*{Divergent terms and the self term in BC}

For the interaction with FF particles in the electrically neutral condition, the $\boldsymbol{k}= \boldsymbol{0}$ term \textcolor{black}{(\ref{eq:12b})} in Eq. (\ref{eq:12}) disappears because of the zero multipole moment $\mathcal{M}_{0}^{0}=0$ for  $Q_{\mathrm{C}}=0$. However, when $Q_{\mathrm{C}}\ne0$, this term is not zero because of non-zero $\mathcal{M}^{0}_{0}\ne 0$. In this case, a big problem arises as shown below, where the term diverges. To avoid this divergence, we introduced the BC such that the total charge of the whole system satisfies electrical neutrality, i.e., $Q_{\mathrm{C}}+\rho _{N+1}V=0$, where $\rho_{N+1}=-Q_{\mathrm{C}}/V$ is the charge density of the BC. 
Then, the electrostatic interaction does not diverge. 

\subsection*{Quadrupole surface term of charged systems}

The divergent $\boldsymbol{k} = \boldsymbol{0}$ term in FMM can be written as the $\boldsymbol{k}$ ${\rightarrow}$ $\boldsymbol{0}$ limit of the reciprocal term (\ref{eq:12b}) by 

\begin{equation}
\lim _{k\rightarrow 0}\frac{\pi ^{\frac{3}{2}}i^{\lambda }}{2^{\lambda - 2}\Gamma \left(\lambda +\frac{1}{2}\right)V}\boldsymbol{\mathcal{I}}_{\lambda }^{\eta }\left(\boldsymbol{k}\right)k^{2\lambda - 1}e^{- \frac{k^{2}}{4\kappa ^{2}}}=\lim _{k\rightarrow 0}\frac{4\pi }{V}\frac{1}{k^{2}}e^{- \frac{k^{2}}{4\kappa ^{2}}}\delta _{\lambda 0}\delta _{\eta 0}, \label{eq:surfcharged1}
\end{equation}
where  zenith, $\theta _{k}$ ,and azimuth angles, $\phi_{k}$ ,of the wave vector $\boldsymbol{k}$ in $\boldsymbol{\mathcal{I}}_{\lambda }^{\eta }\left(\boldsymbol{k}\right)$ are averaged as $\frac{1}{4\pi }\int _{0}^{2\pi }\mathrm{d}\phi _{\boldsymbol{k}}\int _{- 1}^{1}\mathrm{d}\cos \theta _{\boldsymbol{k}} \boldsymbol{\mathcal{I}}_{\lambda }^{\eta }\left(\boldsymbol{k}\right)=\frac{1}{k}\delta _{\lambda 0}\delta _{\eta 0}$.
This term has a non-zero value only when $\lambda =\eta =0$. Then, the potential energy, $V_{\mathrm{PC-PC}}^{\mathrm{M2L},\mathrm{rec}}$, coming from this term in PC-PC interaction listed in Table \ref{TablePCPC} an be written using local expansion coefficient $\boldsymbol{\mathcal{L}}_{\lambda }^{\eta }$ and multipole moment $\boldsymbol{\mathcal{M}}_{\lambda }^{\eta }$  in Eq. (\ref{eq:EqFF}) can be written by
\begin{equation}
\frac{1}{2}\sum _{\lambda =0}^{n_{\max }}\left(- 1\right)^{\lambda }\sum _{\eta =- \lambda }^{\lambda }\boldsymbol{\mathcal{L}}_{\lambda }^{\eta }\boldsymbol{\mathcal{M}}_{\lambda }^{\eta }=\lim _{k\rightarrow 0}\frac{2\pi }{V}\frac{1}{k^{2}}e^{- \frac{k^{2}}{4\kappa ^{2}}}{Q_{\mathrm{C}}}^{2}= \lim _{k\rightarrow 0} V_{\mathrm{PC-PC}}^{\mathrm{M2L},\mathrm{rec}}.  \label{eq:surfcharged2}
\end{equation}
where $n_{\max }$ is the maximum degree of solid harmonics expansion and the multipole moment  $\boldsymbol{\mathcal{M}}_{0}^{0}=Q_{C}$ for PC.

Next, the PC-BC and BC-BC interactions, $V_{\mathrm{PC-BC}}^{\mathrm{M2L},\mathrm{rec}}$ and $V_{\mathrm{BC-BC}}^{\mathrm{M2L},\mathrm{rec}}$, listed in Table \ref{TablePCBC} in far subcells  are difficult to evaluate in FMM using solid harmonics 
In this study, these terms are evaluated using the result of Ewald method, $V_{\mathrm{PC-BC}}^{\mathrm{E},\mathrm{rec}}+V_{\mathrm{BC-BC}}^{\mathrm{E},\mathrm{rec}}$ in Eq. (\ref{eq:surftotal}). Hence, the $\boldsymbol{k} = \boldsymbol{0}$ term of PC-BC and BC-BC interactions in FMM is 
\begin{eqnarray}
\lim _{k\rightarrow 0}\left(V_{\mathrm{PC-BC}}^{\mathrm{M2L},\mathrm{rec}}+V_{\mathrm{BC-BC}}^{\mathrm{M2L},\mathrm{rec}}\right)&=& \lim _{k\rightarrow 0}\left(V_{\mathrm{PC-BC}}^{\mathrm{E},\mathrm{rec}}+V_{\mathrm{BC-BC}}^{\mathrm{E},\mathrm{rec}}\right) \notag \\
&= &\frac{2\pi }{V}\lim _{k\rightarrow 0}\frac{e^{- \frac{k^{2}}{4\kappa ^{2}}}}{k^{2}}\left\{- 2Q_{\mathrm{C}}\sum _{i=1}^{N}q_{i}\frac{\sin \left(kr_{i}\right)}{kr_{i}}+{Q_{\mathrm{C}}}^{2}\right\}.  \label{eq:surfcharged3}
\end{eqnarray}
This term also diverges as in Eq. (\ref{eq:surfcharged2}). Then the sum of the PC-PC, PC-BC, and BC-BC interactions, $i.e.$ the sum of Eqs. (\ref{eq:surfcharged2}) and (\ref{eq:surfcharged3}), is
\begin{equation}
V_{\mathrm{PC-PC}}^{\mathrm{M2L},\mathrm{rec}} +
V_{\mathrm{PC-BC}}^{\mathrm{M2L},\mathrm{rec}}+V_{\mathrm{BC-BC}}^{\mathrm{M2L},\mathrm{rec}}
= \frac{2\pi }{3V}Q_{\mathrm{C}}\sum _{i=1}^{N}q_{i}{r_{i}}^{2} \label{eq:surfchargedend},
\end{equation}
where we used $\sin{(kr)}=kr-\frac{(kr)^3}{3!}+ \mathcal{O}((kr)^5) $. 
This term, including the quadrupole of PC, no longer diverges. 
This quadrupole surface term should be calculated additionally  in FMM for  charged systems.

In summary, electrostatic potential energy of a net charged system consisting of $N$ point charges and  BC using conducting boundary condition at infinite distance can be written by
\begin{eqnarray}
V^{\text{elec}}&=&\frac{1}{2}\sum _{\text{subcell}}\sum _{i\in \text{subcell}}q_{i}\sum _{j\in \mathrm{NF}}\frac{q_{j}}{\left| \boldsymbol{r}_{i}- \boldsymbol{r}_{j}\right| }+\frac{1}{2}\sum _{\text{subcell}}\sum _{i\in \text{subcell}}q_{i}\sum _{n=0}^{n_{\max }}\sum _{m=- n}^{n}\boldsymbol{\mathcal{L}}_{n}^{m}\left(\textit{l }_{\max }\right)\boldsymbol{\mathcal{R}}_{n}^{m}\left(\boldsymbol{r}_{i}- \boldsymbol{r}_{\mathrm{L}'}\right) \notag \\
&-& \frac{\pi }{2V\kappa ^{2}}{Q_{\mathrm{C}}}^{2}+\frac{2\pi }{3V}Q_{\mathrm{C}}\sum _{i=1}^{N}q_{i}{r_{i}}^{2}- \frac{2\pi }{3V}\mathbf{\pmb{\mu} }\cdot \mathbf{\pmb{\mu} },  \label{eq:Velecfull}
\end{eqnarray}
where $\boldsymbol{\mu}$ is the dipole moment in the unit cell,  $\boldsymbol{\mathcal{R}}_{n}^{m}$ is the regular solid harmonics.
The first term on the right side of Eq. (\ref{eq:Velecfull}) is the direct interaction in the NF, the second term is the interaction via a local expansion coefficient from the FF, the third term is the charged system term given in Eq. (\ref{eq:6}), and the fourth and fifth terms are the quadrupole and dipole surface terms, respectively. The first and second terms include the sum over all subcells at $\textit{l}=\textit{l}_{\max }$ in the unit cell. 
For the dipole surface term $\boldsymbol{\mu}$, see Appendix \ref{dipolar} for detail.
It should be noted that $\kappa $ in Eq. (\ref{eq:12})  and $\kappa $ in the third term on the right-hand side of Eq. (\ref{eq:Velecfull}) must be the same value. 

As shown above, the quadrupole surface term was derived naturally when evaluating the \(\boldsymbol{k} = \boldsymbol{0}\) term of the PC-PC, PC-BC, and BC-BC interactions in FMM. Concerning this term, it has been reported that a similar quadrupole term appears in the electrostatic interaction for a net charged system with a periodic boundary condition \cite{redlack1972electrostatic,redlack1975coulombic}. However, it is not obvious how to apply this term when FMM is used in a system with PC and BC. The derivation of Eq. (\ref{eq:Velecfull}) in the present study is the first to provide an electrostatic interaction in FMM for a net charged system.

\subsection*{A new efficient calculation scheme for  solute-solvent interaction energy by FMM}

Evaluation of solute-solvent interaction is  required for free energy calculations such as thermodynamic integration and free energy perturbation methods. Here, an efficient calculation method for the solute-solvent interaction  by FMM is presented.
In the PME method, since the real space term is expressed in the form of two-body interaction, the solute-solvent interaction can be easily extracted. However, the reciprocal space term cannot be expressed by the pair interaction. A special handling is necessary to evaluate the solute-solvent interaction in this term.

Let “u” and “v” be the solute and solvent, respectively. 
Here, total interaction in the reciprocal space and contribution to it from solute-solute and solvent-solvent interactions by the PME method are denoted by $V_{\mathrm{uu},\mathrm{uv},\mathrm{vv}}^{\mathrm{PME},\mathrm{rec}}$, $V_{\mathrm{uu}}^{\mathrm{PME},\mathrm{rec}}$, and $V_{\mathrm{vv}}^{\mathrm{PME},\mathrm{rec}}$, respectively according to the PME method.
Then the solute-solvent interaction can be calculated by
\begin{equation}
V_{\mathrm{uv}}^{\mathrm{PME},\mathrm{rec}}=V_{\mathrm{uu},\mathrm{uv},\mathrm{vv}}^{\mathrm{PME},\mathrm{rec}}- \left(V_{\mathrm{uu}}^{\mathrm{PME},\mathrm{rec}}+V_{\mathrm{vv}}^{\mathrm{PME},\mathrm{rec}}\right). \label{eq:PMEcalc}
\end{equation}
In this case, two interactions $V_{\mathrm{uu}}^{\mathrm{PME},\mathrm{rec}}$ and $V_{\mathrm{vv}}^{\mathrm{PME},\mathrm{rec}}$ have to be calculated in addition to $V_{\mathrm{uu},\mathrm{uv},\mathrm{vv}}^{\mathrm{PME},\mathrm{rec}}$. The total calculation cost is three times higher than the ordinary reciprocal space interaction calculation. Since electrostatic interaction is long-ranged one and, thus, the contribution from the image cells is not negligible for a charged solute molecule, the explicit calculation in reciprocal space is indispensable.

In the FMM, as in the PME method, two-body interactions are calculated with the particles directly in the NF. However, the second term of Eq. (\ref{eq:Velecfull}) presenting the interaction with the particles in the FF cannot be expressed in the form of pair interaction. 
Therefore, similar handling to the case of PME is required to calculate the solute-solvent interactions in the FF
\begin{equation}
V_{\mathrm{uv}}^{\mathrm{FMM},\mathrm{FF}}=V_{\mathrm{uu},\mathrm{uv},\mathrm{vv}}^{\mathrm{FMM},\mathrm{FF}}- \left(V_{\mathrm{uu}}^{\mathrm{FMM},\mathrm{FF}}+V_{\mathrm{vv}}^{\mathrm{FMM},\mathrm{FF}}\right).
\label{eq:FMMconv}
\end{equation}

 Here, let $\Phi _{\mathrm{u},\mathrm{v}}^{\mathrm{FMM},\mathrm{FF}}\left(\boldsymbol{r}\right)$, $\Phi_{\mathrm{u}}^{\mathrm{FMM},\mathrm{FF}}\left(\boldsymbol{r}\right)$, and $\Phi_{\mathrm{v}}^{\mathrm{FMM},\mathrm{FF}}\left(\boldsymbol{r}\right)$ be electric potentials at a position $\boldsymbol{r}$ generated by the solute, solvent, and both solute and solvent molecules, respectively. 
Then, the terms in Eq. (\ref{eq:FMMconv}) may be written by
\begin{eqnarray}
V_{\mathrm{uu},\mathrm{uv},\mathrm{vv}}^{\mathrm{FMM},\mathrm{FF}}&=& \sum _{i\in \mathrm{u}}q_{i} \Phi_{\mathrm{u},\mathrm{v}}^{\mathrm{FMM},\mathrm{FF}}\left(\boldsymbol{r}_{i}\right)+ \sum _{i\in \mathrm{v}}q_{i} \Phi_{\mathrm{u},\mathrm{v}}^{\mathrm{FMM},\mathrm{FF}}\left(\boldsymbol{r}_{i}\right)  \label{Phissswww}\\
V_{\mathrm{uu}}^{\mathrm{FMM},\mathrm{FF}}&=&\sum _{i\in \mathrm{u}}q_{i} \Phi_{\mathrm{u}}^{\mathrm{FMM},\mathrm{FF}}\left(\boldsymbol{r}_{i}\right) \label{Phiss}\\
V_{\mathrm{vv}}^{\mathrm{FMM},\mathrm{FF}}&=&\sum _{i\in \mathrm{v}}q_{i} \Phi_{\mathrm{v}}^{\mathrm{FMM},\mathrm{FF}}\left(\boldsymbol{r}_{i}\right) \label{Phiww},
\end{eqnarray}
where $\Phi_{\mathrm{u}}^{\mathrm{FMM},\mathrm{FF}}\left(\boldsymbol{r}\right)$  can be calculated by the local expansion coefficient $\mathcal{L}_{n~u}^{m}$ generated by the solute molecules as $\Phi_{\mathrm{u}}^{\mathrm{FMM},\mathrm{FF}}\left(\boldsymbol{r}\right)=\sum _{n=0}^{n_{\max }}\sum _{m=- n}^{n}\boldsymbol{\mathcal{L}}^{m}_{n~\mathrm{u}}\boldsymbol{\mathcal{R}}_{n}^{m}\left(\boldsymbol{r}- \boldsymbol{r}_{\mathrm{L}'}\right)$. 
$\Phi_{\mathrm{v}}^{\mathrm{FMM},\mathrm{FF}}\left(\boldsymbol{r}\right)=\sum _{n=0}^{n_{\max }}\sum _{m=- n}^{n}\boldsymbol{\mathcal{L}}^{m}_{n~\mathrm{v}}\boldsymbol{\mathcal{R}}_{n}^{m}\left(\boldsymbol{r}- \boldsymbol{r}_{\mathrm{L}'}\right)$ and
$\Phi_{\mathrm{u},\mathrm{v}}^{\mathrm{FMM},\mathrm{FF}}\left(\boldsymbol{r}\right)=\sum _{n=0}^{n_{\max }}\sum _{m=- n}^{n}\boldsymbol{\mathcal{L}}^{m}_{n~\mathrm{u},\mathrm{v}}\boldsymbol{\mathcal{R}}_{n}^{m}\left(\boldsymbol{r}- \boldsymbol{r}_{\mathrm{L}'}\right)$. The conditions $i\in \mathrm{u}$ and $i\in \mathrm{v}$ are also obtained in  a similar way.
Using Eq. (\ref{eq:FMMconv}), computational cost is still three times higher than the ordinary M2L calculation. This is a bottleneck of the FF calculation. 


Now, we propose a more efficient way to evaluate the solute-solvent interactions. 
The above descriptions, Eqs. (\ref{Phissswww}) - (\ref{Phiww}), provide the following equations
\begin{eqnarray}
\sum _{i\in \mathrm{u}}q_{i}  \Phi_{\mathrm{u},\mathrm{v}}^{\mathrm{FMM},\mathrm{FF}}\left(\boldsymbol{r}_{i}\right) &=& V_{\mathrm{uu}}^{\mathrm{FMM},\mathrm{FF}} + y V_{\mathrm{uv}}^{\mathrm{FMM},\mathrm{FF}} \\
\sum _{i\in \mathrm{u}}q_{i} \Phi_{\mathrm{u}}^{\mathrm{FMM},\mathrm{FF}}\left(\boldsymbol{r}_{i}\right)&=&V_{\mathrm{uu}}^{\mathrm{FMM},\mathrm{FF}} \\
\sum _{i\in \mathrm{v}}q_{i} \Phi_{\mathrm{u}}^{\mathrm{FMM},\mathrm{FF}}\left(\boldsymbol{r}_{i}\right)&=&z V_{\mathrm{uv}}^{\mathrm{FMM},\mathrm{FF}},  
\end{eqnarray}
where y + z =1.
These relations can provide a new efficient solute-solvent interaction calculation scheme as
\begin{eqnarray}
V_{\mathrm{uv}}^{\mathrm{FMM},\mathrm{FF}}&=& \left(
V_{\mathrm{uu}}^{\mathrm{FMM},\mathrm{FF}} + y V_{\mathrm{uv}}^{\mathrm{FMM},\mathrm{FF}} \right) -  \left( V_{\mathrm{uu}}^{\mathrm{FMM},\mathrm{FF}} - z V_{\mathrm{uv}}^{\mathrm{FMM},\mathrm{FF}} \right)\\
&=&
\sum _{i\in \mathrm{u}}q_{i}\Phi_{\mathrm{u},\mathrm{v}}^{\mathrm{FMM},\mathrm{FF}}\left(\boldsymbol{r}_{i}\right)- \left(\sum _{i\in \mathrm{u}}q_{i} \Phi_{\mathrm{u}}^{\mathrm{FMM},\mathrm{FF}}\left(\boldsymbol{r}_{i}\right)- \sum _{i\in \mathrm{v}}q_{i} \Phi_{\mathrm{u}}^{\mathrm{FMM},\mathrm{FF}}\left(\boldsymbol{r}_{i}\right)\right). \label{eq:FMMcalc}
\end{eqnarray}
According to Eq. (\ref{eq:FMMcalc}), calculation of $\Phi_{}^{\mathrm{FMM},\mathrm{FF}}$ must be performed only twice for $\Phi_{\mathrm{u,v}}^{\mathrm{FMM},\mathrm{FF}}$ and $\Phi_{\mathrm{u}}^{\mathrm{FMM},\mathrm{FF}}$. This reduces computational cost at  a bottleneck of FF calculation. 
Alternative equation is also obtained by exchanging subscript u and v in Eq. (\ref{eq:FMMcalc}).

\subsection*{Free energy calculation}

	 We consider calculation of solvation free energy of ionic solutes with net charge \(Q_{\textrm{u}}\)  based on thermodynamic integration  method for a process where the charged solutes in vacuum are immersed into electrolyte solution with excess counter ionic net charge of \(Q_{\textrm{v}}\). 
Though values of \(Q_{\textrm{u}}\) and \(Q_{\textrm{v}}\) are arbitrary from a view point of the present FMM calculation, \(Q_{\textrm{u}}=-Q_{\textrm{v}}\)  in most cases where the resultant solution of interest is electrically neutral. Further, even for the neutral solution, number of ions is still arbitrary. The process is schematically presented in Fig. \ref{ScheNoBG} .

However, in this case, the systems in  Fig. \ref{ScheNoBG} do not satisfy electrically neutral condition so that we cannot apply conventional FMM to these systems. To avoid this, we introduce background charge as shown in Fig. \ref{ScheBG}, where \(Q_{\textrm{u}}+Q_{\textrm{Bu}}=0\), \(Q_{\textrm{v}}+Q_{\textrm{Bv}}=0\), and \(Q_{\textrm{u}}+Q_{\textrm{v}}+Q_{\textrm{Bu}}+Q_{\textrm{Bv}}=0\) . 
\(Q_{\textrm{Bu}}\) and \(Q_{\textrm{Bv}}\) are the background charges introduced to neutralize the solute and solvent charges, respectively. Then, we can evaluate coulombic interactions rigorously using the FMM method presented in this paper for the systems under periodic boundary condition. The reference state of the solvation free energy is a little different from experiment. The ion is not actually in vacuum but is electrically neutralized by the hypothetical background charge. This discrepancy is unavoidable as far as we handle the ions in vacuum. We must be careful when we compare the result with experiment.

The method is useful when we investigate the difference in solvation free energy between two states, say A and B. An interesting example is the free energy of transfer of an  ionic antiviral reagent from outside to inside of the virus capsid. A thermodynamic cycle shown in Fig. \ref{ScheDGAB} gives free energy of transfer \(\Delta G_{\textrm{ AB } } =\Delta G_{\textrm{ B } }-\Delta G_{\textrm{ A } }\) between the two real states. In principle, the difference can be compared with experiment.

In order to transfer the solute from vacuum to solution step by step, the coupling parameter $\lambda $ was introduced into the interaction function as
\begin{eqnarray}
V_{\lambda }&=&V_{\mathrm{u-u}}+\left(1- \lambda \right)V_{\mathrm{u-v}}+V_{\mathrm{v-v}} \notag \\ 
&+&V_{\mathrm{u-Bu}}+\left(1- \lambda \right)V_{\mathrm{u-Bv}}+V_{\mathrm{v-Bv}} +\left(1- \lambda \right)V_{\mathrm{v-Bu}} \notag \\
&+&V_{\mathrm{Bu-Bu}}+\left(1- \lambda \right)V_{\mathrm{Bu-Bv}}+V_{\mathrm{Bv-Bv}} \label{eq:Vswucheme}
.  
\end{eqnarray}
At $\lambda=1$, the solute is in vacuum while it is in the solution at $\lambda=0$.
Thus, the free energy difference between the two thermodynamic states may be calculated by
\begin{equation}
\Delta {G} = \int_{0}^{1}  \left< \frac{\partial V_{\lambda}}{\partial \lambda_{}}  \right>_{\lambda}  
   d \lambda ,  
\end{equation} 
where angle bracket $\left<  \cdots \right>_{\lambda}$ means the ensemble average with the fixed coupling parameter $\lambda$. 
From Eqs. (\ref{eq:Velecfull}), (\ref{eq:FMMcalc}), and (\ref{eq:Vswucheme})  , the total potential energy $V_{\lambda }=V_{\lambda _{\mathrm{C}}}^{\mathrm{FMM}}$ with $\lambda =\lambda _{\mathrm{C}}$ is expressed as
\begin{eqnarray}
V_{\lambda _{\mathrm{C}}}^{\mathrm{FMM}}&=&\sum _{\left(i,j\right)\in \left(\mathrm{u},\mathrm{u}\right),\left(\mathrm{v},\mathrm{v}\right)}^{\mathrm{NF}}\frac{q_{i}q_{j}}{r_{ij}}+\left(1- \lambda _{\mathrm{C}}\right)\sum _{\left(i,j\right)\in \left(\mathrm{u},\mathrm{v}\right)}^{\mathrm{NF}}\frac{q_{i}q_{j}}{r_{ij}} \notag\\
&+&\left(1- \lambda _{\mathrm{C}}\right)\sum _{i\in \mathrm{u}}q_{i}\Phi _{\mathrm{u},\mathrm{v}}^{\mathrm{FMM},\mathrm{FF}}\left(\boldsymbol{r}_{i}\right)+\lambda _{\mathrm{C}}\left\{
\sum _{i\in \mathrm{u}}q_{i} \Phi _{\mathrm{u} \mathrm{}}^{\mathrm{FMM},\mathrm{FF}}\left(\boldsymbol{r}_{i}\right)
- \sum _{i\in \mathrm{v}}q_{i} \Phi _{\mathrm{u} \mathrm{}}^{\mathrm{FMM},\mathrm{FF}}\left(\boldsymbol{r}_{i}\right)
\right\}+\sum _{i\in \mathrm{v}}q_{i}\Phi _{\mathrm{u},\mathrm{v}}^{\mathrm{FMM},\mathrm{FF}}\left(\boldsymbol{r}_{i}\right) \notag \\
&+&\frac{\pi }{V\kappa ^{2}}\left\{\frac{{Q_{\mathrm{C}}}^{2}}{2}- \lambda _{\mathrm{C}}\left(Q_{\mathrm{v}}Q_{\mathrm{Bu}}+Q_{\mathrm{u}}Q_{\mathrm{Bv}}+Q_{\mathrm{Bv}}Q_{\mathrm{Bu}}\right)\right\} \notag \\
&-& \frac{2\pi }{3V}\left\{\left(1- \lambda _{\mathrm{C}}\right)Q_{\mathrm{u}}+Q_{\mathrm{v}}\right\}\sum _{\mathrm{i}\in \mathrm{v}}q_{i}r_{i}^{2}- \frac{2\pi }{3V}\left\{Q_{\mathrm{u}}+\left(1- \lambda _{\mathrm{C}}\right)Q_{\mathrm{v}}\right\}\sum _{\mathrm{i}\in \mathrm{u}}q_{i}r_{i}^{2} \notag \\
&+&\frac{2\pi }{3V}\left(\mu _{\mathrm{u}}^{2}+\mu _{\mathrm{v}}^{2}\right)+\left(1- \lambda _{\mathrm{C}}\right)\frac{4\pi }{3V}\mathbf{\pmb{\mu} }_{\mathrm{u}}\cdot \mathbf{\pmb{\mu} }_{\mathrm{v}}  \label{eq:FMMVTI}
\end{eqnarray}
where the sum in the first term is taken over solute-solute atom pairs and solvent-solvent pairs, and the sum in the second term is taken over solute-solvent atom pairs. Here, $\mathbf{\pmb{\mu }}_{\mathrm{u}}=\sum _{i\in \mathrm{u}}q_{i}\boldsymbol{r}_{i}$, $\mathbf{\pmb{\mu} }_{\mathrm{v}}=\sum _{i\in \mathrm{v}}q_{i}\boldsymbol{r}_{i}$, ${\mu} _{\mathrm{u}}=\left| \mathbf{\pmb{\mu} }_{\mathrm{u}}\right| $, and ${\mu} _{\mathrm{v}}=\left| \mathbf{\pmb{\mu} }_{\mathrm{v}}\right| $.

\subsection*{Calculation detail}

Solvation free energy calculation was performed by program package MODYLAS \cite{andoh2013modylas}  with implementation of the thermodynamic integration algorithm. CHARMM 36 force field \cite{vanommeslaeghe2010charmm,beglov1994finite,klauda2010update} was adopted for Cl$^{-}$, Mg$^{2+}$, and Na$^{+}$ ions. TIP3P model \cite{jorgensen_monte_1985} was used for water. 1080 TIP3P water molecules and a single ion were placed in the cubic unit cell. Electrostatic interaction was calculated by FMM. PME method was also used as a reference calculation. In the FMM calculation, the number of cell length divisions $\textit{l}_{\max }$ was set to \textcolor{black}{3}, and its unit cell was subdivided into \textcolor{black}{8} subgroups in each dimension such that we considered \textcolor{black}{512} finest subcells. The number of expansion of solid harmonics $n_{\textrm{ max }}$ was set to 4.
The splitting parameter $\kappa_{\textrm{ PME } } $ in the PME method was set to be 3.20 nm$^{-1}$, the order of the B-spline interpolation was 4, and the number of FFT grids was set to 64 for each dimension. Cutoff distance of LJ interaction, real space coulombic interaction in the PME method, and the size of the NF space were all 0.8 nm.  Five-fold Nos\'{e}-Hoover chain thermostat \cite{martyna1996explicit} was used for temperature control, where the time constant of the thermostat was $\tau _{Q}=0.5$ ps. Andersen barostat was used for pressure control with the time constant of the barostat $\tau _{W}=0.5$ ps. SHAKE/ROLL and RATTLE/ROLL algorithms were used to constrain the distances between oxygen-hydrogen and hydrogen-hydrogen atoms of water molecules with relative tolerance $1.0\times 10^{-8}$. 
An initial configuration was prepared randomly. After energy minimization by the steepest decent method \cite{andoh2013modylas}, the system was equilibrated by NVT ensemble calculation at temperature \textit{T} = 298.15 K for 50 ps with the time step of $\Delta t=1$ fs. Then, NPT ensemble calculation was performed at temperature \textit{T} = 298.15 K and pressure \textit{P} = 101325.0 Pa for 2 ns with $\Delta t=2$ fs.

 In addition to the coulombic coupling parameter $\lambda_{C}$,  we introduced a coupling parameter $\lambda _{\mathrm{LJ}}$ for  the solute-solvent Lennard-Jones (LJ) interaction. In the case of $\lambda _{\mathrm{LJ}}$ simply applied to the ordinary LJ potential between solute and solvent such as linear interpolation, abrupt overlapping of atoms may occur near $\lambda _{\mathrm{LJ}}=0$, i.e.,  in the generating process, which may cause divergence of energy. 
To avoid this, the LJ potential was modified using "soft-core" potential \cite{zacharias_separationshifted_1994} as 
\begin{equation}
V_{\lambda _{\mathrm{LJ}}}^{\mathrm{LJ}}=\left(1- \lambda _{\mathrm{LJ}}\right)\sum _{\left(i,j\right)\in \left(\mathrm{u},\mathrm{v}\right)}4\varepsilon _{ij}\left[\left\{\frac{A_{ij}}{\left(\alpha _{\mathrm{LJ}}\lambda _{\mathrm{LJ}}{A_{ij}}^{c}+{r_{ij}}^{c}\right)^{\frac{1}{c}}}\right\}^{12}- \left\{\frac{B_{ij}}{\left(\alpha _{\mathrm{LJ}}\lambda _{\mathrm{LJ}}{B_{ij}}^{c}+{r_{ij}}^{c}\right)^{\frac{1}{c}}}\right\}^{6}\right]
\end{equation}
where, $\varepsilon _{ij}$, $A_{ij}$, and $B_{ij}$ are parameters of the LJ interaction, and $\alpha _{\mathrm{LJ}}$ and \textit{c} are newly introduced parameters to avoid divergence. In this study, the parameters were set to be $\alpha _{\mathrm{LJ}}=0.2$ and \textit{c}= 6. It should be noted
that this soft-core approach is also applicable to electrostatic NF term. 

Thus, for the thermodynamic integration calculation, we adopted a two-step TI path with two coupling parameters, $\lambda _{\mathrm{C}}$ and $\lambda _{\mathrm{LJ}}$, where solute-solvent coulombic and LJ interactions decrease step by step separately. The values of the parameters were  $\lambda _{\mathrm{C}}=(0.00,0.02,0.04,0.06,\ldots ,0.98,1.00)$ at $\lambda _{\mathrm{LJ}}=0.00$, and subsequently $\lambda _{\mathrm{LJ}}=(0.00,0.02,0.04,0.06,\ldots ,0.98,1.00)$ at $\lambda _{\mathrm{C}}=1.00$. 
Thus, the path was divided into 101 steps, which gave sufficiently smooth change of the state. 
The MD calculations were performed both by FMM and PME method. Initial configurations of these MD calculations were independent of each other. 
 The total hydration free energy was obtained by the sum of the numerical integration of the first derivative of potential energy function with respect to $\lambda _{\mathrm{C}}$ and $\lambda _{\mathrm{LJ}}$. 

\section*{Results and discussion}

\subsection*{Electrostatic potential energy}

To demonstrate the correctness of the new expression of FMM, total electrostatic potential energy of the whole system as well as the contribution  to it from  solute-solvent interactions 
was calculated both by FMM and PME method. The target system was a solution in which a single Mg$^{2+}$ ion is dissolved in water. Figure \ref{errorFMMPME} shows the relative error 
\begin{equation}
\text{relative}\,\text{error}=\left| \frac{V^{\mathrm{FMM}}- V^{\mathrm{PME}}}{V^{\mathrm{PME}}}\right|
\end{equation}
in the calculated electrostatic potential energy of the whole system and the contribution from the solute-solvent interactions at $\left(\lambda_\mathrm{CL},\lambda _{\mathrm{LJ}}\right)=\left(0.0,0.0\right)$. It is defined by the difference in the calculated coulombic potential energy obtained by FMM relative to that  by  high-accuracy PME calculation. The error was evaluated  as a function of degree of expansion $n_{\textrm{max}}$ of solid harmonics using 100 configurations from the trajectory. 
Figure \ref{errorFMMPME} clearly shows that the logarithm of the relative error decreases almost linearly as a function of the order of expansion of solid harmonics. The error in the coulombic potential energy of the whole system is less than $10^{-5}$ and $10^{-6}$ at $n_{\max }=4$  and $6$, respectively. 
The error is small enough for most of the MD calculations.
This means that the present expression of FMM for the charged system electrically neutralized by the  BC is correct, and that can be controlled by $n_{\max }$. However, because there is a trade-off relationship between computational cost and accuracy,  we must adopt best $n_{\max }$. 

Here, in our present expression, FMM employs the conducting boundary condition in Eq. (\ref{eq:Velecfull}), where a quadrupole surface term represented by Eq. (\ref{eq:surfchargedend}) is taken into account in the case of the system with BC. 
Without this quadrupole term, a discontinuity is produced in the electrostatic interaction of the system when a charged molecule crosses the unit cell boundary. This causes instability of the MD calculation. In this sense, the new term presented in this study  is essential in the MD calculation by FMM for the system with BC.

\subsection*{Solvation free energy of ions}

In order to demonstrate validity of our formulation for the free energy calculations, we evaluated the free energy of three kinds of ion (Mg$^{2+}$, Na$^{+}$, Cl$^{-}$) in pure water by FMM and compared it  with the one by PME method. Here, one solute  ion, which was first in vacuum, was immersed into pure water. 
In this case, $Q_{\textrm{u  } }= +2e, +e,$ and $-e$ for Mg$^{2+}$, Na$^{+}$, and Cl$^{-}$, respectively, and $Q_{\textrm{v  } }=0$, $Q_{\textrm{ Bu } }=-Q_{\textrm{u  } }$, and $Q_{\textrm{ Bv } }=0$, according to the notation shown in Figs. \ref{ScheNoBG} and \ref{ScheBG}.

A typical example of cumulative average of the integrand $\frac{\partial V_{\lambda}}{\partial \lambda_{\textrm{ C } }}$ of the present thermodynamic integration calculation for Mg$^{2+}$ at $\left(\lambda_\mathrm{C},\lambda _{\mathrm{LJ}}\right)=\left(0.36, 0.00\right)$ is shown in Fig \ref{Mg_cumu}. 
The calculated instantaneous values are also plotted. 
The average of both FMM and PME calculations converged well within 1.0 ns. 

The averaged integrand $\left\langle \frac{\mathrm{d}V_{\lambda_{\mathrm{C}}}^{\mathrm{}}}{\mathrm{d}\lambda_\mathrm{C}}\right\rangle $  for \textcolor{black}{Mg}$^{2+}$ \textcolor{black}{ion} is shown in Fig. \ref{Mg_average_all} as a function of $\lambda_\mathrm{C}$. The results of FMM and PME method agree very well with each other. The solvation free energy of the ion calculated by FMM and PME methods by numerically integrating $\left\langle \frac{\mathrm{d}V_{\lambda _{\mathrm{C}}}^{\mathrm{}}}{\mathrm{d}\lambda_\mathrm{C}}\right\rangle $ and $\left\langle \frac{\mathrm{d}V_{\lambda _{\mathrm{LJ}}}^{\mathrm{}}}{\mathrm{d}\lambda _{\mathrm{LJ}}}\right\rangle $ over $\lambda_\mathrm{C}$ and $\lambda _{\mathrm{LJ}}$ is listed in Table \ref{tblSFE}. The two solvation free energies are in excellent agreement with each other.

\section*{Conclusion}

A new expression has been derived for long-ranged electrostatic interactions among PC and BC based upon FMM as shown Eqs. (\ref{eq:12}) and (\ref{eq:Velecfull}). In the expression, three new terms were taken into account in the lattice sum over the image cells, i.e., the charged system term in Eq. (\ref{eq:6}), the quadrupole  surface term due to the BC in Eq. (\ref{eq:surfcharged3}), and the self term (\ref{eq:12c}).

Further, an efficient calculation scheme for solute-solvent interactions by FMM was also proposed in Eq. (\ref{eq:FMMcalc}), which is required in thermodynamic integration calculation and free energy perturbation calculation.
These formulas were applied to evaluate the solvation free energy of Mg\(^{2+}\), Na\(^{+}\), and Cl\(^{{-}}\) ions in water. 
The solvation free energy obtained by the FMM was in excellent agreement with that of the reference calculations by the PME method. The present  efficient calculation of solute-solvent evaluations may be applied to other methods such as multistate Bennett acceptance ratio (MBAR) and weighted histogram analysis method (WHAM).

The present method gives a new and efficient way  of free energy calculation of ionic solutes in large systems with more than ten million atoms,  such as binding free energy of a protein and hydration free energy of a large charged colloid in electrolyte solution.

\section*{Acknowledgment}

This work was done by the support of FLAGSHIP2020, MEXT within the priority issues 1 (Development of Next-Generation Drug Design Technology) and 5 (Development of new fundamental technologies for high-efficiency energy creation, conversion/storage and use) using computational resources of the K computer provided by the RIKEN Advanced Institute for Computational Science through the HPCI System Research project (Project ID: hp150164, hp150269, hp150275). Calculations were partly performed at the Information Technology Center of Nagoya University, at the Institute for Solid State Physics, the University of Tokyo, and at the Research Center for Computational Science, Okazaki, Japan, and the Supercomputer Center. This work was also funded by JSPS KAKENHI Grant Number JP17K04758 (N.Y.). We thank to Dr. Sakashita and Dr. Andoh for helping implementation of our method into MODYLAS program.

\section*{AIP Publishing Data Sharing Policy}

The data that support the findings of this study are available from the corresponding author upon reasonable request.

\appendix

\section{ Ewald summation for the system with PC and BC under conducting boundary condition}

\label{ewaldcharge}

We outline the Ewald summation of a system consisting of point charges (PC) and background charge (BC).

Let an atom \textit{i} with a point charge $q_{i}$ be at a position $\boldsymbol{r}_{i}$. We suppose that there are \textit{N} charged atoms in a cubic unit cell with a side length \textit{b}  and volume $V=b^{3}$. Periodic boundary condition is imposed on the unit cell. Now, it is assumed that the net charge $Q_{\mathrm{C}}=\sum _{i=1}^{N}q_{i}$ of PC is not zero. Then, to satisfy the electrical neutrality, a uniform background charge (BC)  density is introduced in the system. The charge density of the BC is $\rho _{N+1}=- \frac{Q_{\mathrm{C}}}{V}$, where BC may be regarded as the (\textit{N}  + 1)-th charge, i.e. $q_{N+1}=- Q_{\mathrm{C}}$.  Then, the charge density of the system can be expressed as
\begin{equation}
\rho \left(\boldsymbol{r}\right)=\sum _{i=1}^{N}q_{i}\left\{\delta \left(\boldsymbol{r}- \boldsymbol{r}_{i}\right)- \frac{1}{V}\right\},	
\end{equation}
using Dirac's delta function $\delta \left(\boldsymbol{r}\right)$. Hereafter, we use Gaussian units and omit the factor $\frac{1}{4\pi \varepsilon _{0}}$ in the expression of the electrostatic interaction energy just for simplicity, where $\varepsilon _{0}$ is the dielectric constant of vacuum. The electrostatic interaction of the system consisting of PC and BC is expressed by
\begin{eqnarray}
V^{\text{elec}}&=&\frac{1}{2}\sum _{\mathbf{\pmb{\nu} }}'\int _{V}\int _{V}\frac{\rho \left(\boldsymbol{r}\right)\rho \left(\boldsymbol{r}'\right)}{\left| \boldsymbol{r}- \boldsymbol{r}'+\mathbf{\pmb{\nu} }b\right| }\mathrm{d}\boldsymbol{r}\mathrm{d}\boldsymbol{r}' \notag \\
&=&\frac{1}{2}\sum _{\mathbf{\pmb{\nu} }}'\sum _{i=1}^{N}\sum _{j=1}^{N}\frac{q_{i}q_{j}}{\left| \boldsymbol{r}_{i}- \boldsymbol{r}_{j}+\mathbf{\pmb{\nu} }b\right| }- \frac{1}{V}\sum _{\mathbf{\pmb{\nu} }}\sum _{i=1}^{N}\sum _{j=1}^{N}\int _{V}\frac{q_{i}q_{j}}{\left| \boldsymbol{r}_{i}- \boldsymbol{r}'+\mathbf{\pmb{\nu} }b\right| }\mathrm{d}\boldsymbol{r}' \notag \\
&+&\frac{1}{2V^{2}}\sum _{\mathbf{\pmb{\nu} }}\sum _{i=1}^{N}\sum _{j=1}^{N}\int _{V}\int _{V}\frac{q_{i}q_{j}}{\left| \boldsymbol{r}- \boldsymbol{r}'+\mathbf{\pmb{\nu} }b\right| }\mathrm{d}\boldsymbol{r}\mathrm{d}\boldsymbol{r}' \label{eq:Velec}
\end{eqnarray}
where $\mathbf{\pmb{\nu} }=\left(\begin{array}{l}
\nu _{x}\\
\nu _{y}\\
\nu _{z}
\end{array}\right)$ 
is a vector consisting of a set of three integers. $\sum _{\mathbf{\pmb{\nu} }}'$ means that at $\mathbf{\pmb{\nu} }=\mathbf{0}$, $i=j$ is excluded from the sum. It should be noted that the BC-BC interaction is included in the electrostatic interaction using $\rho_{N+1}=-\sum_{i=1}^{N}q_{i}/V $. 

The Ewald method is applied to Eq. (\ref{eq:Velec}). The potential energy function under conducting boundary condition is then \cite{frenkel2001understanding}
\begin{eqnarray}
V^{\text{elec}}&=&
\frac{1}{2}\sum _{\mathbf{\pmb{\nu} }}'\sum _{i=1}^{N}\sum _{j=1}^{N}\frac{q_{i}q_{j}\text{erfc}\left(\kappa \left| \boldsymbol{r}_{i}- \boldsymbol{r}_{i}+\mathbf{\pmb{\nu} }b\right| \right)}{\left| \boldsymbol{r}_{i}- \boldsymbol{r}_{i}+\mathbf{\pmb{\nu} }b\right| } +
\frac{2\pi }{V}\sum _{\mathbf{k}\neq \mathbf{0}}\frac{e^{- \frac{k^{2}}{4\kappa ^{2}}}}{k^{2}}\sum _{l=1}^{N}q_{l}e^{i\mathbf{k}\cdot {\boldsymbol{r}_{l}}}\sum _{j=1}^{N}q_{j}e^{- i\mathbf{k}\cdot {\boldsymbol{r}_{j}}}  \notag \\
&-& \frac{\kappa }{\sqrt{\pi }}\sum _{i=1}^{N}q_{i}^{2} 
- \frac{\pi }{2V\kappa ^{2}}{Q_{\mathrm{C}}}^{2}, 
\end{eqnarray}
where the first term is the real space term, the second term is the reciprocal space term, the third term is the self term, and the fourth term is the charged system term. Further, a dipole surface term, 
\begin{equation}
\frac{2\pi }{3V}\mathbf{\pmb{\mu} }\cdot \mathbf{\pmb{\mu} }, \label{surfdipole}
\end{equation}
may be added if the vacuum boundary condition is employed at the infinite boundary. 
The derivation of each term is given below in detail. 

In molecular dynamics simulation, the surface term should not be included because this term causes discontinuity of energy when a charged particle crosses the wall of box. Hence, we need to remove the surface term. The condition is, then, equal to the conducting boundary condition.

In the Ewald method, each term in $V^{\text{elec}}$ is separated into the real space terms and the reciprocal space terms starting from $\frac{1}{r}=\frac{\text{erfc}\left(\kappa r\right)}{r}+\frac{\mathrm{erf}\left(\kappa r\right)}{r}$ using a complementary error function $\text{erfc}\left(\kappa r\right)$ and an error function $\mathrm{erf}\left(\kappa r\right)$ with the Ewald splitting parameter $\kappa $. 
Concerning the PC-PC interaction, i.e., the first term of the second equation in Eq. (\ref{eq:Velec}), the real space term is 
\begin{equation}
V_{\mathrm{PC-PC}}^{\mathrm{E},\text{real}}=\frac{1}{2}\sum _{\mathbf{\pmb{\nu} }}'\sum _{i=1}^{N}\sum _{j=1}^{N}\frac{q_{i}q_{j}\text{erfc}\left(\kappa \left| \boldsymbol{r}_{i}- \boldsymbol{r}_{i}+\mathbf{\pmb{\nu} }b\right| \right)}{\left| \boldsymbol{r}_{i}- \boldsymbol{r}_{i}+\mathbf{\pmb{\nu} }b\right| }, \label{eq:Ereal}
\end{equation}
which may be  evaluated directly in the real space.

The reciprocal space term of the PC-PC interaction is obtained by Fourier transformation of the complementary error function of the PC-PC interaction. However, the term of \textit{i} = \textit{j} at $\mathbf{\pmb{\nu} }=\mathbf{0}$  is missing. By adding this term and performing Fourier transform, 
\begin{equation}
V_{\mathrm{PC-PC}}^{\mathrm{E},\mathrm{rec}}=\frac{2\pi }{V}\sum _{\mathbf{k}\neq \mathbf{0}}\frac{e^{- \frac{k^{2}}{4\kappa ^{2}}}}{k^{2}}\sum _{l=1}^{N}q_{l}e^{i\mathbf{k}\cdot {\boldsymbol{r}_{l}}}\sum _{j=1}^{N}q_{j}e^{- i\mathbf{k}\cdot {\boldsymbol{r}_{j}}} \label{eq:VPPrec}
\end{equation}
is obtained for $\boldsymbol{k} {\neq} \boldsymbol{0}$, where $\boldsymbol{k}$  is a reciprocal lattice vector and $k=\left| \boldsymbol{k}\right| $. Note that the contribution of \textit{i} = \textit{j} at $\mathbf{\pmb{\nu} }=\mathbf{0}$ added in the derivation of Eq. (\ref{eq:VPPrec}) must be subtracted. This self term can be expressed as \cite{frenkel2001understanding}
\begin{equation}
V_{\mathrm{PC-PC}}^{\mathrm{E},\text{self}}=- \frac{\kappa }{\sqrt{\pi }}\sum _{i=1}^{N}q_{i}^{2}. \label{eq:Eself}
\end{equation}

Contribution of PC-BC and BC-BC from the real space, called “charged system term” $V^{\mathrm{cha}}$, can be evaluated by \cite{hummer1996free}
\begin{equation}
V_{\mathrm{PC-BC}}^{\mathrm{E},\text{real}}+V_{\mathrm{BC-BC}}^{\mathrm{E},\text{real}}=V^{\mathrm{cha}}=- \frac{\pi }{2V\kappa ^{2}}{Q_{\mathrm{C}}}^{2}. \label{eq:6}
\end{equation}

The reciprocal space terms PC-BC and BC-BC for $\boldsymbol{k} \neq \boldsymbol{0}$ do not contribute to the potential energy because the Fourier transform of the uniform charge distribution of BC is zero. 

The $\boldsymbol{k} = \boldsymbol{0}$ term of PC-PC, PC-BC, and BC-BC can be rewritten by
\begin{eqnarray}
&& \lim _{k\rightarrow 0}\left(V_{\mathrm{PC-PC}}^{\mathrm{E},\mathrm{rec}}+V_{\mathrm{PC-BC}}^{\mathrm{E},\mathrm{rec}}+V_{\mathrm{BC-BC}}^{\mathrm{E},\mathrm{rec}}\right) \notag  \\
&=&\frac{2\pi }{V}\lim _{k\rightarrow 0}\frac{e^{- \frac{k^{2}}{4\kappa ^{2}}}}{k^{2}}\left\{\sum _{i=1}^{N}q_{i}e^{i\boldsymbol{k}\cdot {\boldsymbol{r}_{i}}}\sum _{j=1}^{N}q_{j}e^{- i\boldsymbol{k}\cdot {\boldsymbol{r}_{j}}}+\rho _{N+1}V\left(\sum _{i=1}^{N}q_{i}e^{i\boldsymbol{k}\cdot {\boldsymbol{r}_{i}}}+\sum _{j=1}^{N}q_{j}e^{- i\boldsymbol{k}\cdot {\boldsymbol{r}_{j}}}\right)+\left(\rho _{N+1}V\right)^{2}\right\}  \notag \\
&=&\frac{2\pi }{V}\lim _{k\rightarrow 0}\frac{e^{- \frac{k^{2}}{4\kappa ^{2}}}}{k^{2}}\left\{\sum _{i=1}^{N}\sum _{j=1}^{N}q_{i}q_{j}\frac{\sin \left(kr_{ij}\right)}{kr_{ij}}- 2Q_{\mathrm{C}}\sum _{i=1}^{N}q_{i}\frac{\sin \left(kr_{i}\right)}{kr_{i}}+{Q_{\mathrm{C}}}^{2}\right\} \notag \\
&=&\frac{2\pi }{3V}\boldsymbol{\pmb{\mu} }\cdot \boldsymbol{\pmb{\mu} }, \label{eq:surftotal}
\end{eqnarray}
where $r_{ij}=\left| \boldsymbol{r}_{i}- \boldsymbol{r}_{j}\right| $ and $\boldsymbol{\pmb{\mu} }=\sum _{i=1}^{N}q_{i}\boldsymbol{r}_{i}$. In the second equation, $e^{\pm i\boldsymbol{k}\cdot \boldsymbol{r}}$ is averaged over a solid angle of $\boldsymbol{k}$, $\frac{1}{4\pi }\int _{0}^{2\pi }\mathrm{d}\phi _{\boldsymbol{k}}\int _{- 1}^{1}\mathrm{d}\cos \theta _{\boldsymbol{k}}e^{\pm i\boldsymbol{k}\cdot \boldsymbol{r}}=\frac{\sin \left(kr\right)}{kr}$. The final formula in Eq. (\ref{eq:surftotal}) is called “surface term”, $V^{\text{surf}}$, and has been discussed in relation to boundary condition of image cells at infinity \cite{frenkel2001understanding,redlack1975coulombic,de1980simulation,ballenegger2009simulations,hu_infinite_2014}. Here, since the conducting boundary condition is adopted, this term is zero. However, if the vacuum boundary condition is chosen, the surface term should be included in the electrostatic interaction. Summation of Eqs. (\ref{eq:Ereal})-(\ref{eq:6}) is the interaction obtained by the Ewald method with the conducting boundary condition  (Table \ref{Table10}). 

The terms $V_{\mathrm{PC-BC}}^{\mathrm{E},\mathrm{rec}}$ and $V_{\mathrm{BC-BC}}^{\mathrm{E},\mathrm{rec}}$ in Eq. (\ref{eq:surftotal}) are also used in the evaluation of PC-BC and BC-BC term in FMM for Eq. (\ref{eq:surfcharged3}).

\section{ Conventional FMM}

\label{APPDConvFMM}
\subsection*{Division of cells}

\begin{figure}[htbp]

\includegraphics[width=0.7\textwidth]{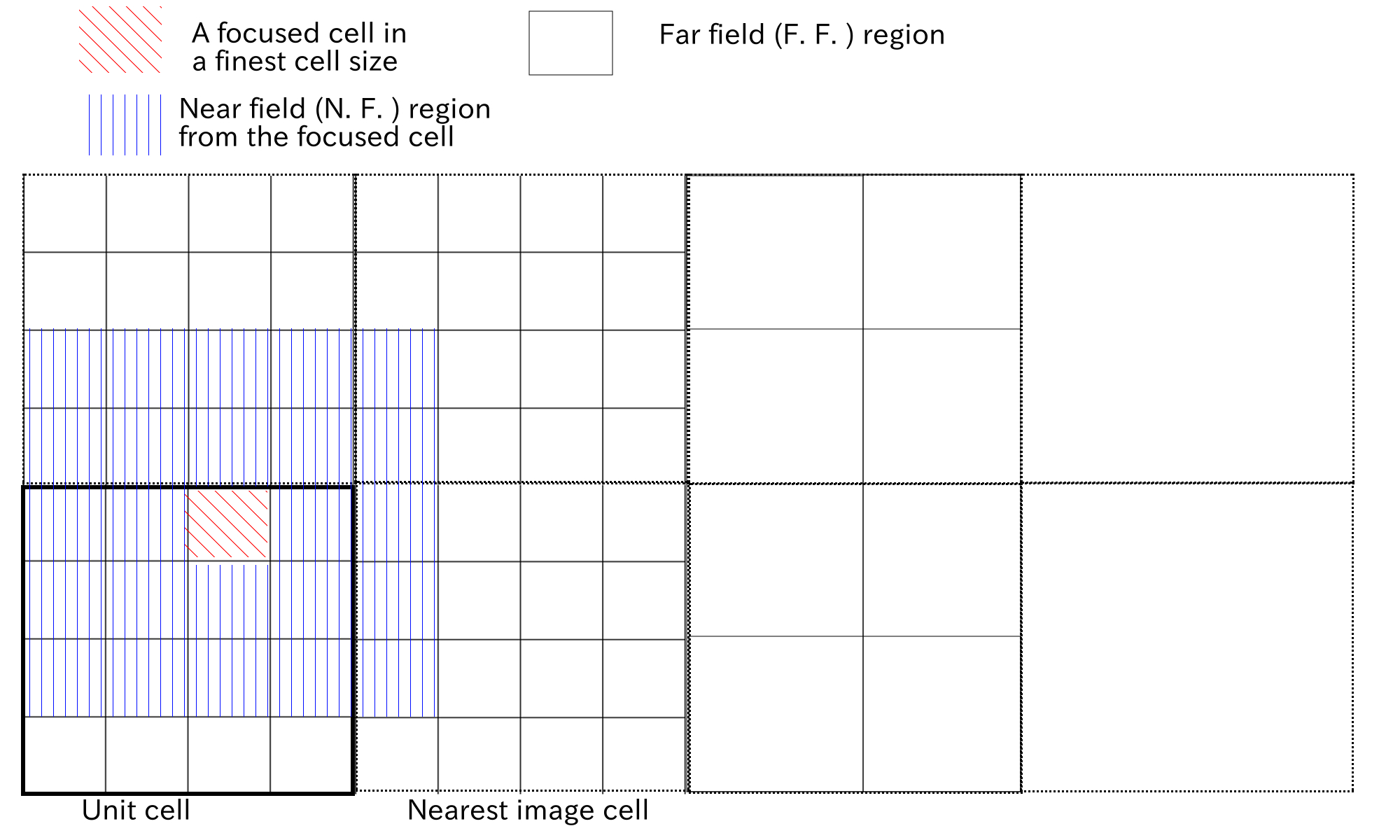}
\caption{\label{Fig1}
Two-dimensional schematic view of the cell division of unit cell and image cells. Interaction of any atom in a focused cell (red) chosen from the finest (\(\textit{l}_{\textrm{max}}=2\)) subcells is separated into near filed (blue) and far field (white) interactions depending on the distance from the focused subcell.}
\end{figure}

\begin{figure}[htbp]

\includegraphics[width=0.7\textwidth]{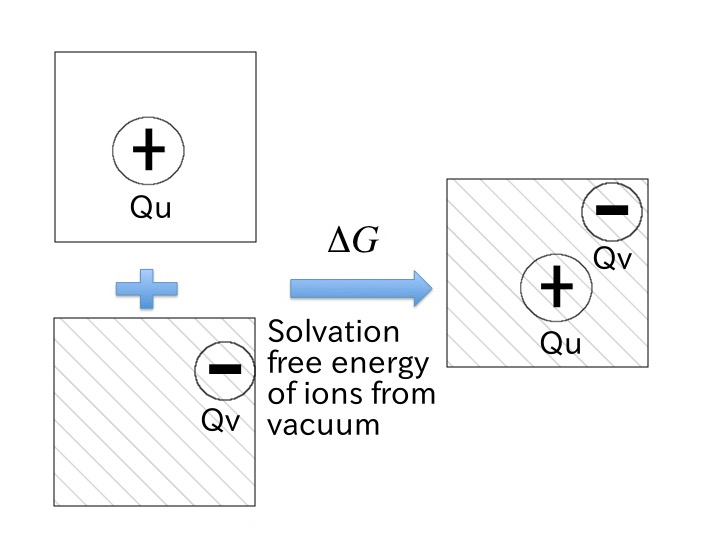}
\caption{\label{ScheNoBG}
Schematic view of a solvation process of an ion from vacuum and the relevant solvation free energy. Stripes represent solvent molecules. Charge neutrality is not satisfied in vacuum or in solution for the states in the left hand side.}
\end{figure}

\begin{figure}[htbp]

\includegraphics[width=0.7\textwidth]{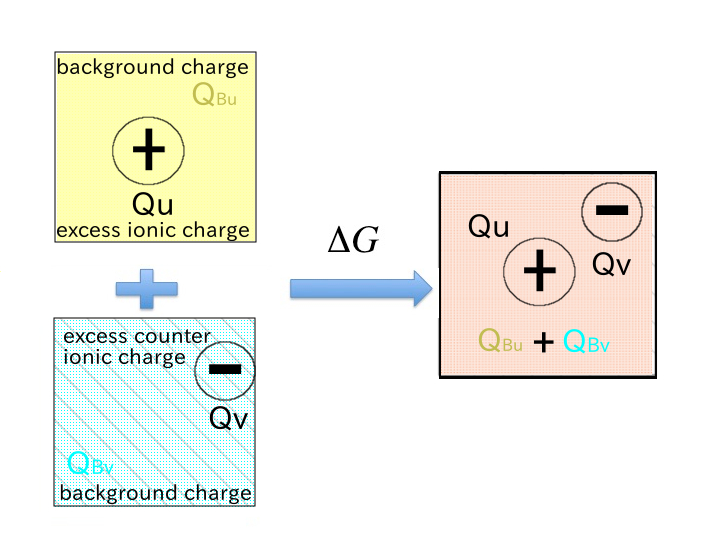}
\caption{\label{ScheBG}
Schematic view of a solvation process of an ion from vacuum and the relevant solvation free energy for the system with background charge density (B.C.), where charge neutrality can be satisfied in vacuum and in solution for the states both in the left hand side and right hand side. Stripes represent solvent molecules and colored backgrounds represent the background charges.}
\end{figure}

\begin{figure}[htbp]

\includegraphics[width=0.7\textwidth]{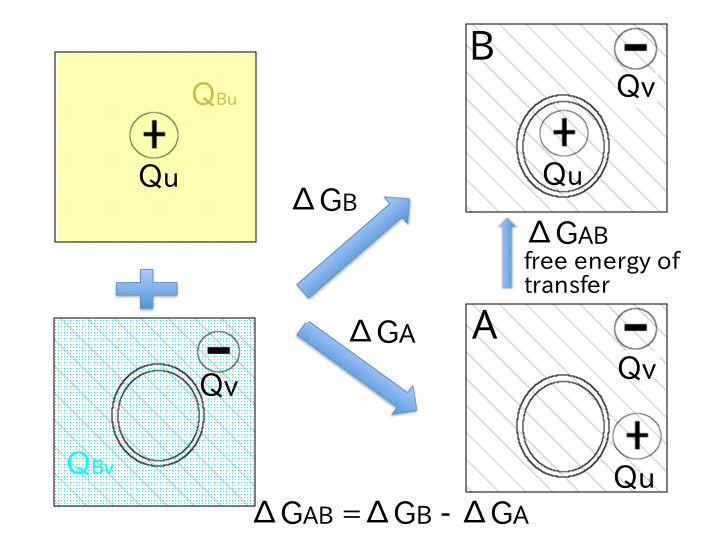}
\caption{\label{ScheDGAB}
Schematic view of a thermodynamic cycle for free energy of transfer of an ion from state A to state B. In the state A, the ion is located outside the capsid, while, in the state B, it is inside the capsid. Stripes and colors are the same as in Fig. \ref{ScheNoBG}}
\end{figure}

\begin{figure}[htbp]

\includegraphics[width=0.7\textwidth]{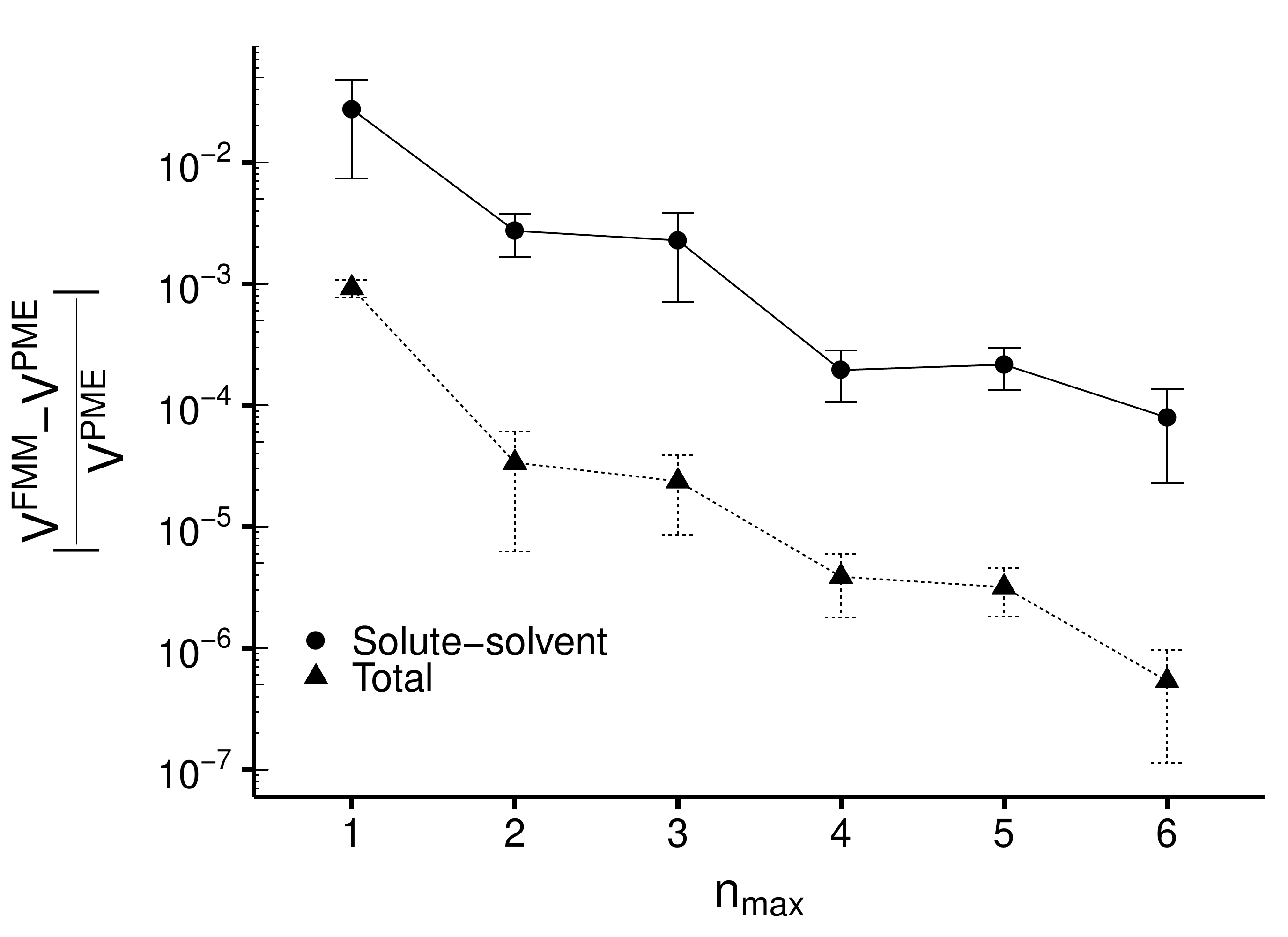}
\caption{\label{errorFMMPME}
Relative error of coulombic potential energy calculated by FMM  from a reference PME calculation with background charge density as a function of  \(n_{\max}\) averaged over 100 independent configurations.}
\end{figure}

\begin{figure}[htbp]

\includegraphics[width=1.0\textwidth]{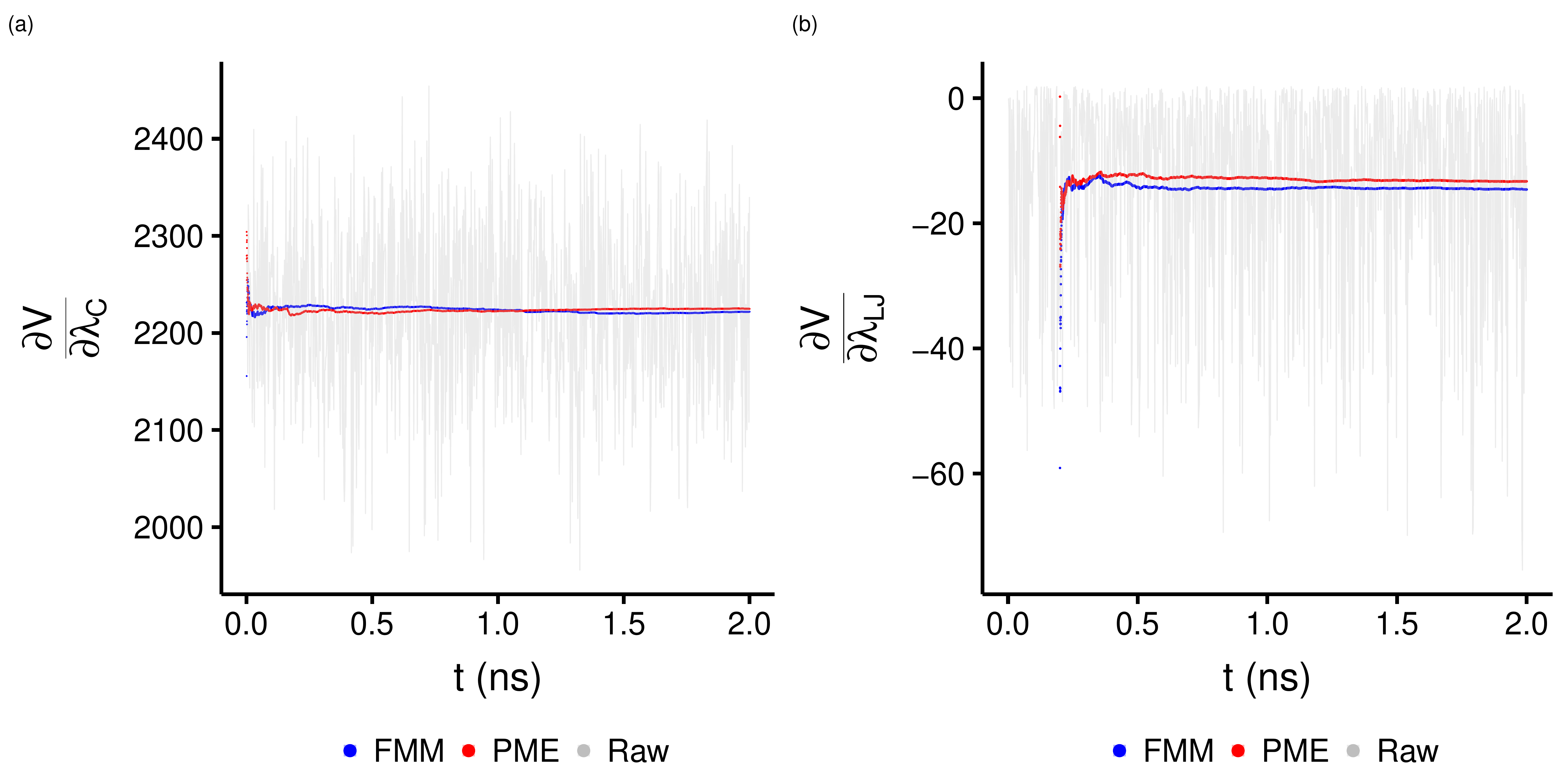}
\caption{\label{Mg_cumu}
Instantaneous value of \(\frac{\mathrm{d}V}{\mathrm{d}\lambda}\) and  its cumulative average calculated by FMM (blue) and PME (red) for the derivative with respect to  the coulombic (left) and LJ (right) coupling parameters for Mg\(^{2+}\).  Circles represent cumulative average and lines are instantaneous values by the FMM.}
\end{figure}

\begin{figure}[htbp]

\includegraphics[width=0.7\textwidth]{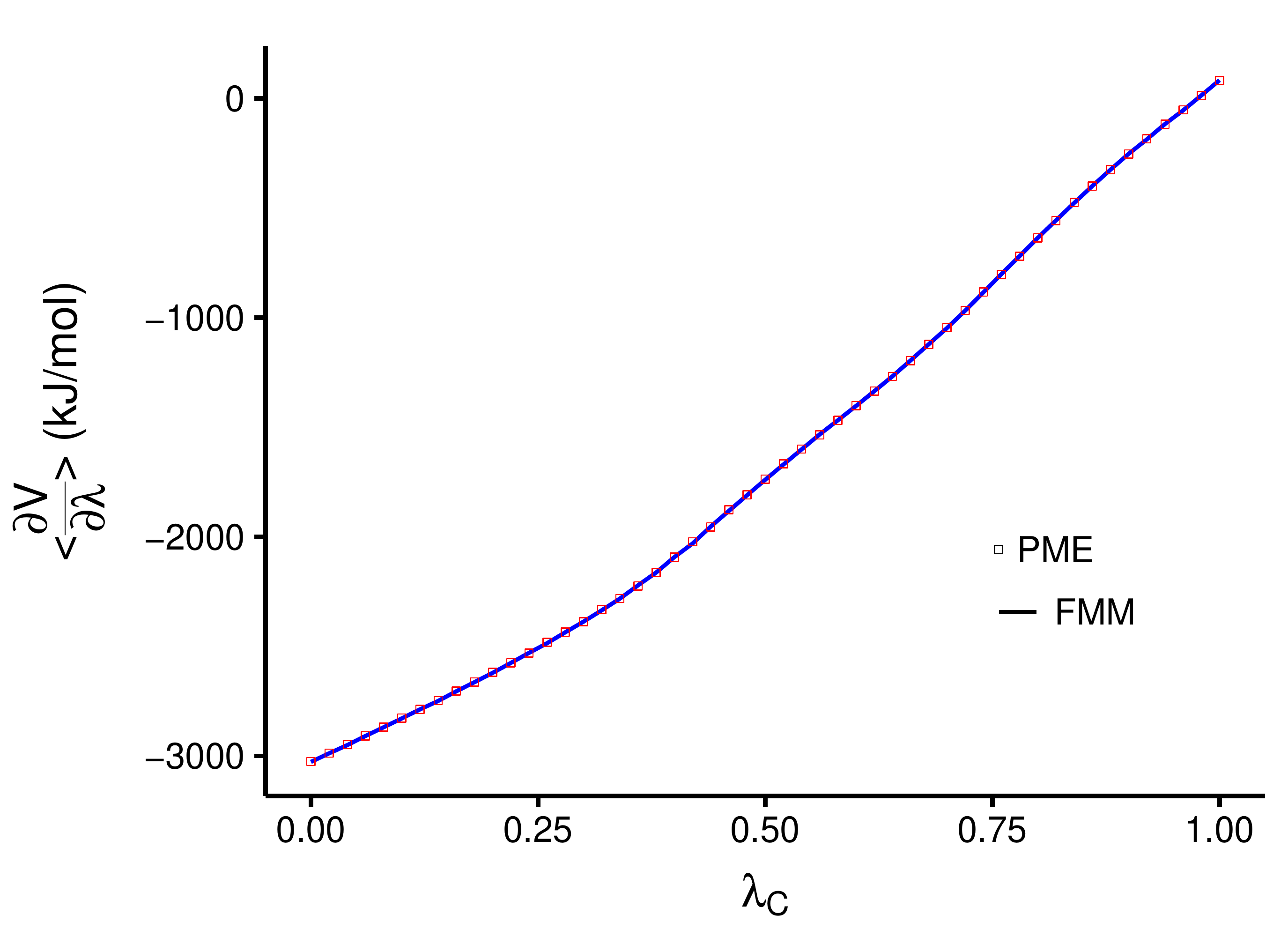}
\caption{\label{Mg_average_all}
The calculated \(\frac{\mathrm{d}V}{\mathrm{d}\lambda}\)  with coulombic couplings by FMM (lines) and PME(square) for the removal of Mg\(^{2+}\) from the solution.}
\end{figure}

In the conventional FMM, a unit cell is divided into hierarchical smaller subcells recursively, as shown in Fig. \ref{Fig1} for two-dimensional case. 
For subcell layers, the octree structure is used, in which each side of a large subcell is divided into half lengths. 
Thus, the octree structure with $\mathit{l }$-th division has $8^{\mathit{l }}$ subcells in three dimensional simulation. 
Here, we employ the $\mathit{l }_{\max }$ as the maximum number of divisions and $0$-th division as a single unit cell. 
In Fig. \ref{Fig1}, the division of the unit cell and first and second image cells at $\mathit{l }_{\max }=2$ is shown. 
According to the FMM, the calculation method of the electrostatic interaction depends on a distance from a cell  of interest to the cells separated into the near-field and far-field shown in red in Fig. \ref{Fig1}. 
The electrostatic interaction  is then evaluated as a sum of the contributions from the NF and FF.

\subsection*{Interaction with charges in far-field}

Interaction with the charges outside the NF is calculated by the local expansion in far subcells in white including image cells as shown in Fig. \ref{Fig1}.
 We show briefly how to calculate PC-PC interaction with charges in FF using the solid harmonics. In the finest subcells at $\textit{l }=\textit{l }_{\max }$, each multipole moment at the center of each subcell is calculated by

\begin{equation}
\boldsymbol{\mathcal{M}}_{n}^{m}\left( l_{\max }\right)=\sum _{j}q_{j}\mathcal{R}_{n}^{m}\left(\boldsymbol{r}_{\mathrm{M}}- \boldsymbol{ r }_{j}\right),  \label{eq:P2M}
\end{equation}
where ${\boldsymbol{ r}}_{\mathrm{M}}$ is the center of the subcell, and $\boldsymbol{\mathcal{R}}_{n}^{m}\left(\boldsymbol{{ r}}\right)$ is the regular solid harmonics with degree $n$ and order $m$. The maximum value of $n$ is $n_{\max }$, and the range of $n$ is $0\leq n\leq n_{\max }$ \cite{nitadori_particle_2014}. The sum of \textit{j} is taken over all PCs in the subcell. This process is called “P2M”. 

Next, a multipole moment in coarser subcells at $\textit{l }=\textit{l }_{\max }- 1$ is calculated by integrating neighboring subcells at $\textit{l }=\textit{l }_{\max }$ with translational shift of the central position of multipole moments. This process (M2M) is repeated until $\boldsymbol{\mathcal{M}}_{n}^{m}\left(\textit{l }=0\right)$ is obtained. These multipole moments are converted to a local expansion coefficient (M2L) which is a Taylor expansion with an arbitrary origin $\boldsymbol{ r }_{\mathrm{L}}$. The local expansion coefficient is used to obtain the electric field far from the center of multipole moments and is written as 

\begin{equation}
\overline{\boldsymbol{\mathcal{L}}}_{n}^{m}=\sum _{\lambda =0}^{n_{\max }}\sum _{\eta =- \lambda }^{\lambda }\boldsymbol{\mathcal{M}}_{\lambda }^{\eta }\boldsymbol{\mathcal{I}}_{n+\lambda }^{- \left(m+\eta \right)}\left(\boldsymbol{ r}_{\mathrm{L}}- \boldsymbol{ r}_{\mathrm{M}}\right), \label{eq:Lcoeffdef}
\end{equation}
where $\boldsymbol{\mathcal{I}}_{n}^{m}\left(\boldsymbol{{ r}}\right)$ is the irregular solid harmonics with degree $n$ and order $m$. 
$\overline{\boldsymbol{\mathcal{L}}}_{n}^{m}$ is the local expansion coefficient. 
The bar on the top means that this local expansion coefficient is obtained from M2L. 
This conversion from multipole moment to local expansion coefficient is done for all subcells in the interaction list in which adjacent subcells to be transformed are listed \cite{greengard1988rapid}. 
 This process is repeated for all $\mathit{{{l } } }$ layers. 
In particular, the contribution from all image cells at $\mathit{{{l } } }=0$ is evaluated using lattice sum, stated in Eq. (\ref{eq:12} )  \cite{schmidt1991implementing,amisaki_precise_2000,zhou1995new}. 
In this case, the irregular solid harmonics $\boldsymbol{\mathcal{I}}_{n+\lambda }^{- \left(m+\eta \right)}\left(\boldsymbol{r}_{\mathrm{L}}- \boldsymbol{r}_{\mathrm{M}}\right)$ in Eq. (\ref{eq:Lcoeffdef}) may be replaced by $\sum _{\boldsymbol{\nu }\neq \boldsymbol{0}}\boldsymbol{\mathcal{I}}_{n+\lambda }^{- \left(m+\eta \right)}\left(\boldsymbol{ r}_{\mathrm{L}}- \boldsymbol{r}_{\boldsymbol{\nu }}\right)$. 
 By multiplying the multipole moment of the unit cell to this summed value, the contribution from the image cell can be incorporated into the local expansion coefficient at ${\textit{{l } } }=0$. 

Once we obtain $\overline{\boldsymbol{\mathcal{L}}}_{n}^{m}\left(0\right)$ at $\textit{l }=0$, we transform it into $\underline{\boldsymbol{\mathcal{L}}}_{n}^{m}\left(1\right)$ by L2L transform where the bar on the bottom means that this local expansion coefficient is obtained from L2L. Then, $\underline{\boldsymbol{\mathcal{L}}}_{n}^{m}\left(1\right)$ and $\overline{\boldsymbol{\mathcal{L}}}_{n}^{m}\left(1\right)$ are merged into one, and we obtain $\boldsymbol{\mathcal{L}}_{n}^{m}\left(1\right)=\overline{\boldsymbol{\mathcal{L}}}_{n}^{m}\left(1\right)+\underline{\boldsymbol{\mathcal{L}}}_{n}^{m}\left(1\right)$ at $\textit{l }=1$. This process is repeated until $\textit{l }=\textit{l }_{\max }$. Finally, the potential energy of the subcell at $\textit{l }=\textit{l }_{\max }$ from the FF used for free energy calculation is evaluated from $\boldsymbol{\mathcal{L}}_{n}^{m}\left(\textit{l }_{\max }\right)$ by
\begin{eqnarray}
V^{\mathrm{FMM},\mathrm{FF}}_{\mathrm{PC-PC}}&=&\frac{1}{2}\sum _{i\in \text{subcell}}q_{i}\sum _{n=0}^{n_{\max }}\sum _{m=- n}^{n}\boldsymbol{\mathcal{L}}_{n}^{m}\left(\textit{l }_{\max }\right)\boldsymbol{\mathcal{R}}_{n}^{m}\left(\boldsymbol{ r }_{i}- \boldsymbol{r }_{\mathrm{L}'}\right) \notag \\
&=&\frac{1}{2}\sum _{n=0}^{n_{\max }}\left(- 1\right)^{n}\sum _{m=- n}^{n}\boldsymbol{\mathcal{L}}_{n}^{m}\left(\textit{l}_{\max }\right)\boldsymbol{\mathcal{M}}_{n}^{m}\left(\textit{l }_{\max }\right), \label{eq:EqFF}
\end{eqnarray}
where $\boldsymbol{r }_{\mathrm{L}'}$ is the origin of the subcell. 

\subsection*{Interaction with charges in near-field}

 The electrostatic interaction of an atom $i$ in a red subcell with all atoms \textit{j} in the same subcell and in the subcells within the second nearest neighbors at finest level $\textit{l }=\textit{l }_{\max }$ shown in blue (near-field:NF) is calculated by the direct PC-PC coulombic interaction. Then, the interaction of the subcell with NF is
\begin{equation}
V^{\mathrm{FMM},\mathrm{NF}}_{\mathrm{PC-PC}}=\frac{1}{2}\sum _{i\in \text{subcell}}q_{i}\sum _{j\in \mathrm{NF}}\frac{q_{j}}{r_{ij}}.      \label{eq:EqNF}
\end{equation}

\section{ Self term in the lattice sum of FMM}

\label{selfterm}

In charge-neutral systems, the self term vanishes so that the self term is not written in papers. Here, the self term is specified, including its derivation. 
In M2L transformation under periodic boundary conditions, the contribution from the image cells was divided into the incomplete gamma function and its complementary function. The reciprocal space term can be derived by Fourier transform of the term including the incomplete gamma function. Then, the missing $\mathbf{\pmb{\nu} }=\mathbf{0}$ term is added. It is necessary to subtract this contribution again. 
An analytical form of the $\mathbf{\pmb{\nu} }=\mathbf{0}$ term is given by replacing the complementary function in the right-hand side of Eq. (\ref{eq:12}) with an incomplete gamma function and taking the limit $r_{\nu }\rightarrow 0$ as
\begin{equation}
\lim _{r_{\nu }\rightarrow 0}\boldsymbol{\mathcal{I}}_{\lambda }^{\eta }\left(\mathbf{r}_{\mathbf{\pmb{\nu} }}\right)\frac{\gamma \left(\lambda +\frac{1}{2},\kappa ^{2}{r_{\mathbf{\pmb{\nu} }}}^{2}\right)}{\Gamma \left(\lambda +\frac{1}{2}\right)}.	 \label{eq:A1}
\end{equation}

The limit of $r_{\nu }\rightarrow 0$ does not depend on the direction of $\boldsymbol{r}_{\mathbf{\pmb{\nu} }}$. Therefore, the solid harmonics is averaged over solid angle of $\boldsymbol{r}_{\mathbf{\pmb{\nu} }}$. Then, due to the orthogonality of solid harmonics, only $\frac{1}{r_{\nu }}\delta _{\lambda 0}\delta _{\eta 0}$ remains. The incomplete gamma function in Eq. (\ref{eq:A1}) may be written by the integral form
\begin{equation}
\gamma \left(\alpha ,x\right)=\int _{0}^{x}e^{- t}t^{\alpha - 1}\mathrm{d}t=\beta ^{\alpha }\int _{0}^{x/\beta }e^{- \beta t}t^{\alpha - 1}\mathrm{d}t. \label{eq:A2} 
\end{equation}
By substituting $x=\kappa ^{2}{r_{\mathbf{\pmb{\nu} }}}^{2}$, $\beta =r_{\nu }^{2}$, and $\alpha =\lambda +\frac{1}{2}$, and inserting Eq. (\ref{eq:A2}) to Eq. (\ref{eq:A1}), it becomes
\begin{equation}
\frac{2\kappa }{\sqrt{\pi }}\delta _{\lambda 0}\delta _{\eta 0}. 	
\end{equation}

\section{ Dipole surface term and infinite boundary condition}

\label{dipolar}

In FMM, it is known that the default infinite boundary condition is the vacuum boundary condition, which includes dipole surface term \cite{schwegler1997linear,herce_electrostatic_2007,zacharias_separationshifted_1994}. Therefore, in order to realize the conducting boundary condition for a stable molecular dynamics simulation by FMM, the dipole surface term represented by Eq. (\ref{surfdipole}) in Appendix \ref{ewaldcharge} should be subtracted from the electrostatic interaction with FMM.
Thus, we obtain the dipole surface term $- \frac{2\pi }{3V}\mathbf{\pmb{\mu} }\cdot \mathbf{\pmb{\mu} }$ 
of the FMM to realize the conducting boundary condition.

\section*{}

\clearpage
\newpage

\begin{table}
\caption{Expressions used in the interaction calculation of a system consisting of PC and BC with periodic boundary conditions.  }
\label{Table2}
\begin{subtable}{\linewidth}
\caption{Formula for calculating PC-PC interaction.}
\label{TablePCPC}
\begin{tabularx}{\textwidth}{|
p{\dimexpr 0.16\linewidth-2\tabcolsep}|
P{\dimexpr 0.2\linewidth-2\tabcolsep}|
P{\dimexpr 0.08\linewidth-2\tabcolsep}|
P{\dimexpr 0.17\linewidth-2\tabcolsep}|
P{\dimexpr 0.39\linewidth-2\tabcolsep}|}
\cline{1-5}
\multicolumn{4}{|l|}{\textbf{FMM}} & \textbf{PC-PC} \\
\cline{1-5}
\multicolumn{4}{|l|}{\textbf{near field}} & Eq. (\ref{eq:EqNF}) \\
\cline{1-5}
\multirow{4}{*}{\textbf{far field}}  & \multicolumn{3}{c|}{\textbf{unit cell}  \par \textbf{near neighbor image cells}} & Eq. (\ref{eq:EqFF}) \\
\cline{2-5}
 & \multirow{3}{*}{\textbf{image cells}}  & \multicolumn{2}{c|}{\textbf{real}} & Eq. (\ref{eq:EqFF}) via term (\ref{eq:12a}) \\
\cline{3-5}
 &  & \multirow{2}{*}{\textbf{rec}}  & \textbf{\textit{k}}\textbf{${\neq}$0} & Eq. (\ref{eq:EqFF}) via term (\ref{eq:12b}) \par  and term (\ref{eq:12c}) \\
\cline{4-5}
 &  &  & \textbf{\textit{k}}=\textbf{0} & Eq. (\ref{eq:surfcharged2}) \\ 
\cline{1-5}
\end{tabularx}
 \end{subtable}

 \begin{subtable}{\linewidth}
\caption{Formula for calculating PC-BC and BC-BC interactions.}
\label{TablePCBC}
\begin{tabularx}{\textwidth}{|
p{\dimexpr 0.16\linewidth-2\tabcolsep}|
P{\dimexpr 0.21\linewidth-2\tabcolsep}|
P{\dimexpr 0.28\linewidth-2\tabcolsep}|
P{\dimexpr 0.36\linewidth-2\tabcolsep}|}
\cline{1-4}
\multicolumn{2}{|l|}{\textbf{Ewald}} & \textbf{PC-BC} & \textbf{BC-BC} \\
\cline{1-4}
\multicolumn{2}{|l|}{\textbf{real}} & \multicolumn{2}{c|}{Eq. (\ref{eq:6})} \\
\cline{1-4}
\multirow{2}{*}{\textbf{rec}}  & \textbf{\textit{k}}\textbf{${\neq}$0} & \textbf{0} & \textbf{0} \\
\cline{2-4}
 & \textbf{\textit{k}}\textbf{=0} & \multicolumn{2}{c|}{Eq. (\ref{eq:surfcharged3})} \\
\cline{1-4}
\end{tabularx}
\end{subtable}
\end{table}

\begin{table*}[htbp]
\caption{\label{tblSFE}
The calculated solvation free energy of ions in water \((kJ/mol)\). Error was calculated by bootstrap sampling.}
\centering
\begin{tabular}{ccc}
\hline
ion & PME & FMM\\
\hline
Cl\(^{-}\) & -342.2 \(\pm\) 0.6 & -342.3 \(\pm\) 0.6\\
\hline
Mg\(^{2+}\) & -1642 \(\pm\) 3 & -1642 \(\pm\) 3\\
\hline
Na\(^{+}\) & -333.8  \(\pm\) 0.7 & -333.8 \(\pm\) 0.7\\
\hline
\end{tabular}
\end{table*}

\setcounter{table}{0}
\renewcommand{\thetable}{A-\arabic{table}}  
\begin{table}
\caption{Expressions used in the Ewald method of a system consisting of particle charge (PC) and background charge (BC) under periodic boundary conditions. }
\label{Table10}
\begin{tabularx}{\textwidth}{|
P{\dimexpr 0.11\linewidth-2\tabcolsep} |
P{\dimexpr 0.22\linewidth-2\tabcolsep}|
P{\dimexpr 0.24\linewidth-2\tabcolsep}|
P{\dimexpr 0.19\linewidth-2\tabcolsep}|
P{\dimexpr 0.24\linewidth-2\tabcolsep}|}
\cline{1-5} \multicolumn{2}{|c|}{\textbf{Ewald}} & \textbf{PC-PC} & \textbf{PC-BC} & \textbf{BC-BC}  \\
\cline{1-5} \multicolumn{2}{|c|}{\textbf{real}} & Eq. (\ref{eq:Ereal}) & \multicolumn{2}{c|}{Eq. (\ref{eq:6})} \\
\cline{1-5}  \multirow{2}{*}{\textbf{rec}}  & \textbf{\textit{k}}\textbf{${\neq}$0} & Eqs. (\ref{eq:VPPrec}) + (\ref{eq:Eself})  & 0 & 0  \\
\cline{2-5}  & \textbf{\textit{k}}\textbf{=0} & \multicolumn{3}{c|}{\multirow{2}{*}{ \shortstack{Eq. (\ref{eq:surftotal}) ~~~(vacuum) \\  0 ~~~(conducting)}}} \\
& & \multicolumn{3}{l|}{}\\
\cline{1-5} 
\end{tabularx}
\end{table}

\clearpage

\listoftables
\listoffigures

\bibliographystyle{apsrev4-1}
\bibliography{fmmbg}

\clearpage
\newpage




\section*{}

\subsection*{Supporting infomation : Exact electrostatic energy calculation for charged systems neutralized by uniformly distributed background charge using fast multipole method and its application to efficient free energy calculation}

\subsection*{Ryo Urano, Wataru Shinoda, Noriyuki Yoshii, Susumu Okazaki}

\addcontentsline{toc}{chapter}{Second unnumbered chapter}
\setcounter{section}{0}
\setcounter{figure}{0}
\setcounter{table}{0}
\setcounter{equation}{0}

\renewcommand{\thepage}{S\arabic{page}} 
\renewcommand{\thesection}{S\arabic{section}}  
\renewcommand{\thetable}{S\arabic{table}}  
\renewcommand{\thefigure}{S\arabic{figure}} 
\renewcommand{\theequation}{S\arabic{equation}} 


\subsection*{\(\frac{\mathrm{d}V}{\mathrm{d}\lambda}\) analysis for Na\(^{+}\) system and Cl\(^{-}\) system}

The same analysis for average \(\frac{\mathrm{d}V}{\mathrm{d}\lambda}\) values in Na and Cl system as that in Mg system are shown. These values are in better agreement between FMM and PME than those of Mg system.

\begin{figure}[htbp]

\includegraphics[width=1.0\textwidth]{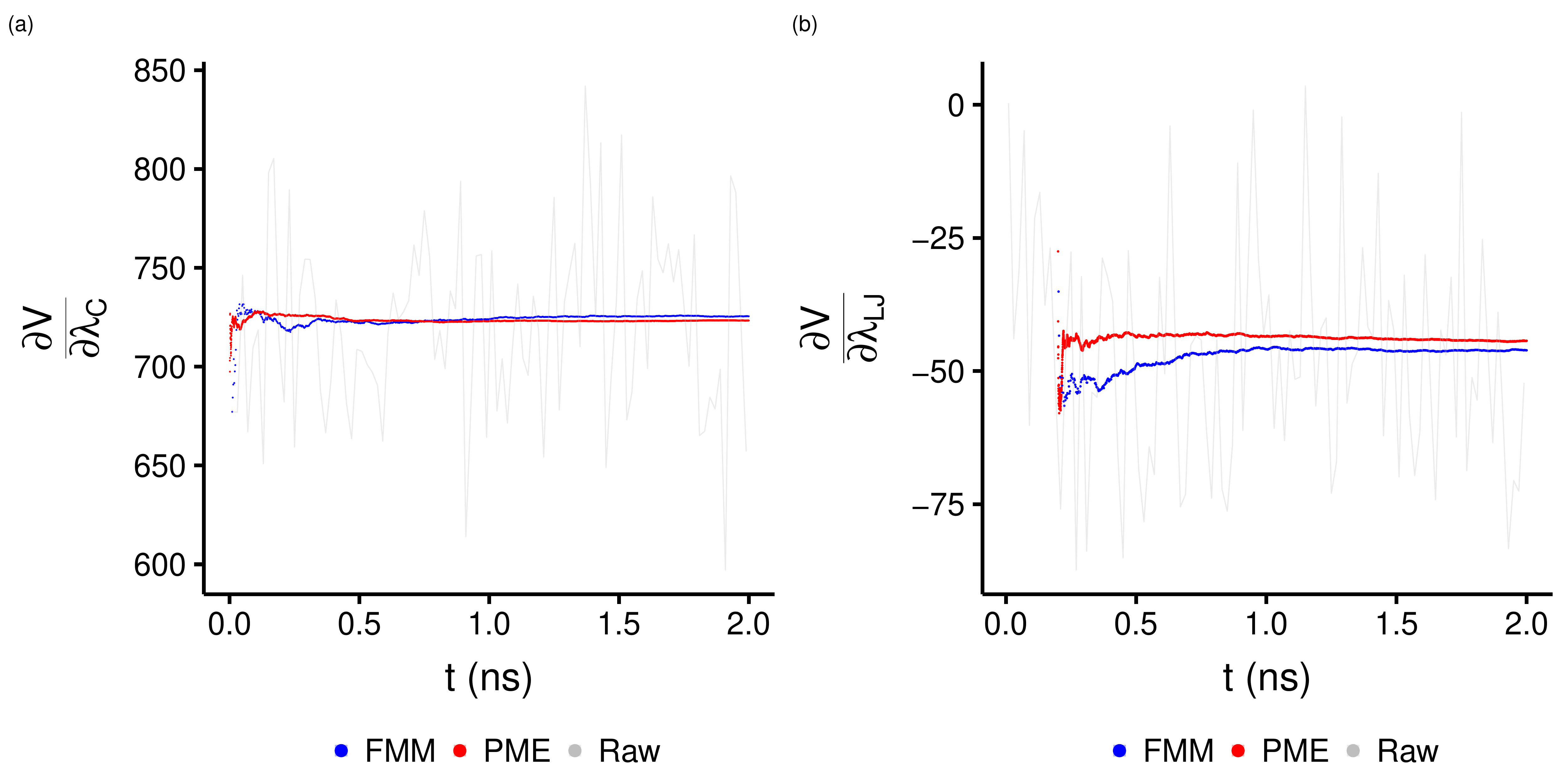}
\caption{\label{cumu}
Instantaneous value of \(\frac{\mathrm{d}V}{\mathrm{d}\lambda}\) and  its cumulative average calculated by FMM (blue) and PME (red) for the derivative with respect to  the coulombic (left) and LJ (right) coupling parameters for Cl\(^{-}\).  Circles represent cumulative average and lines are instantaneous values by the FMM.}
\end{figure}

\begin{figure}[htbp]

\includegraphics[width=0.7\textwidth]{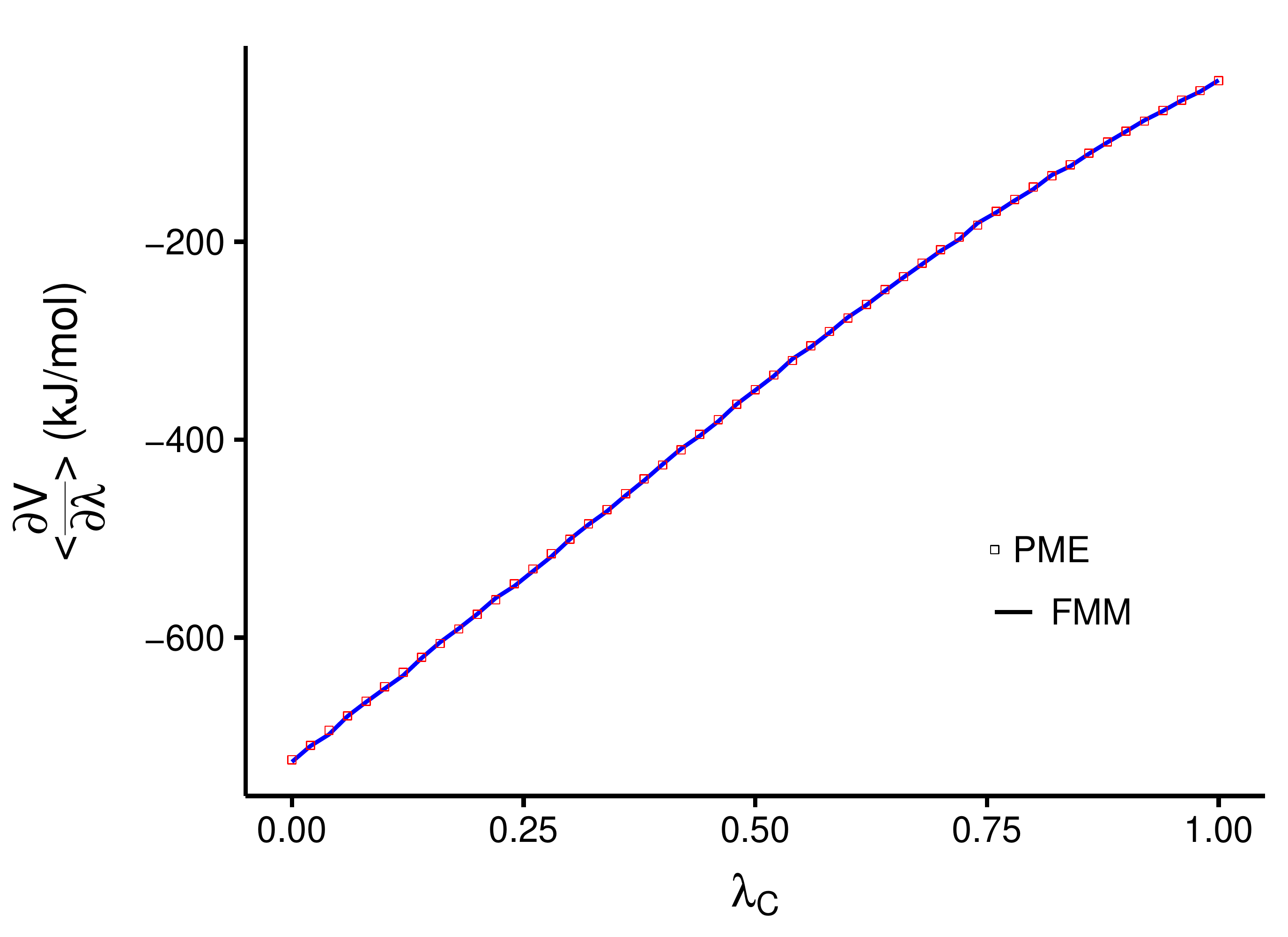}
\caption{\label{average_all}
The calculated \(\frac{\mathrm{d}V}{\mathrm{d}\lambda}\)  with coulombic couplings by FMM (lines) and PME(square) for the removal of Cl\(^{-}\) from the solution.}
\end{figure}

\begin{figure}[htbp]

\includegraphics[width=0.8\textwidth]{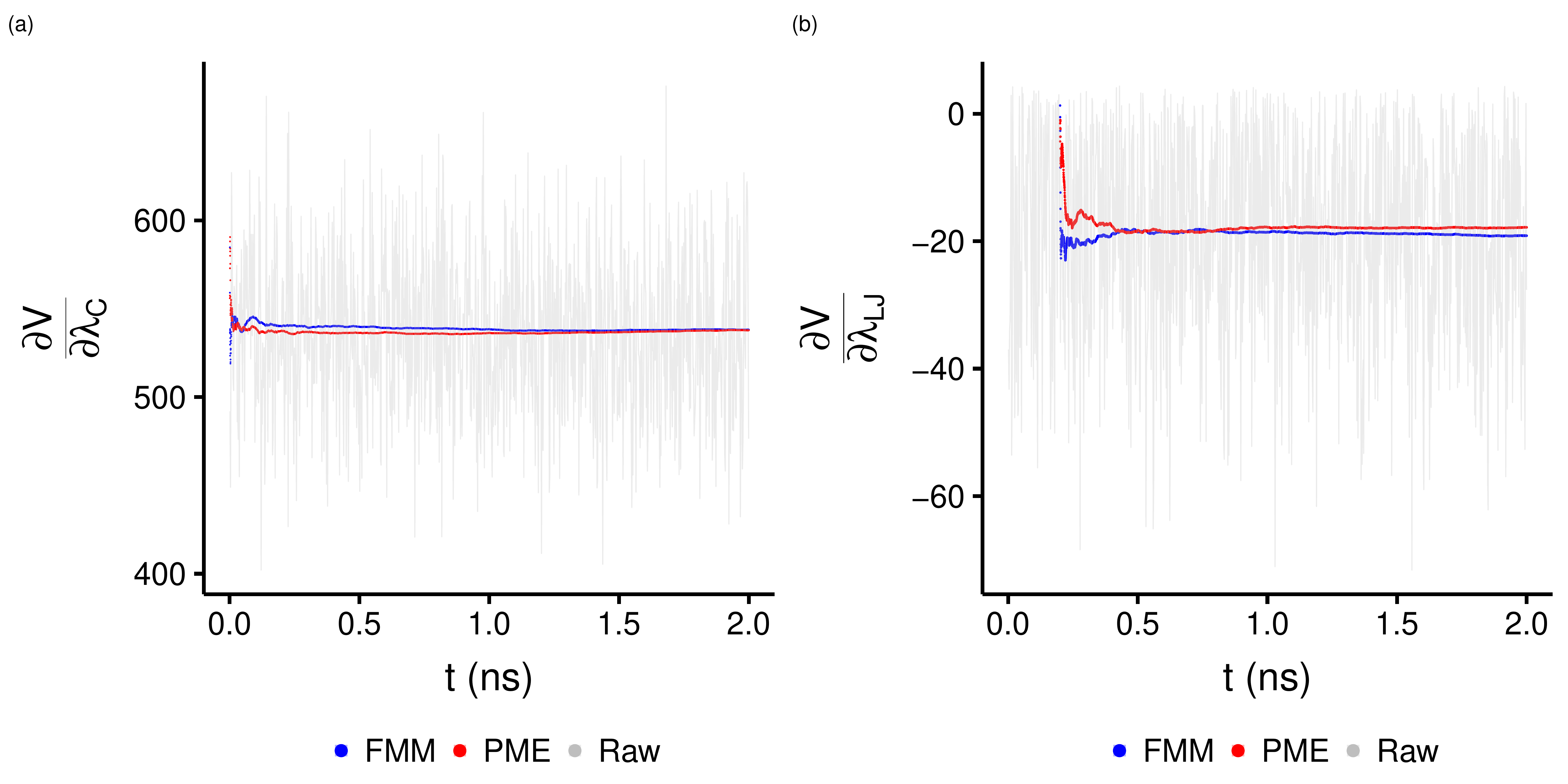}
\caption{\label{Na_cumu}
Instantaneous value of \(\frac{\mathrm{d}V}{\mathrm{d}\lambda}\) and  its cumulative average calculated by FMM (blue) and PME (red) for the derivative with respect to  the coulombic (left) and LJ (right) coupling parameters for Na\(^{+}\).  Circles represent cumulative average and lines are instantaneous values by the FMM.}
\end{figure}

\begin{figure}[htbp]

\includegraphics[width=0.7\textwidth]{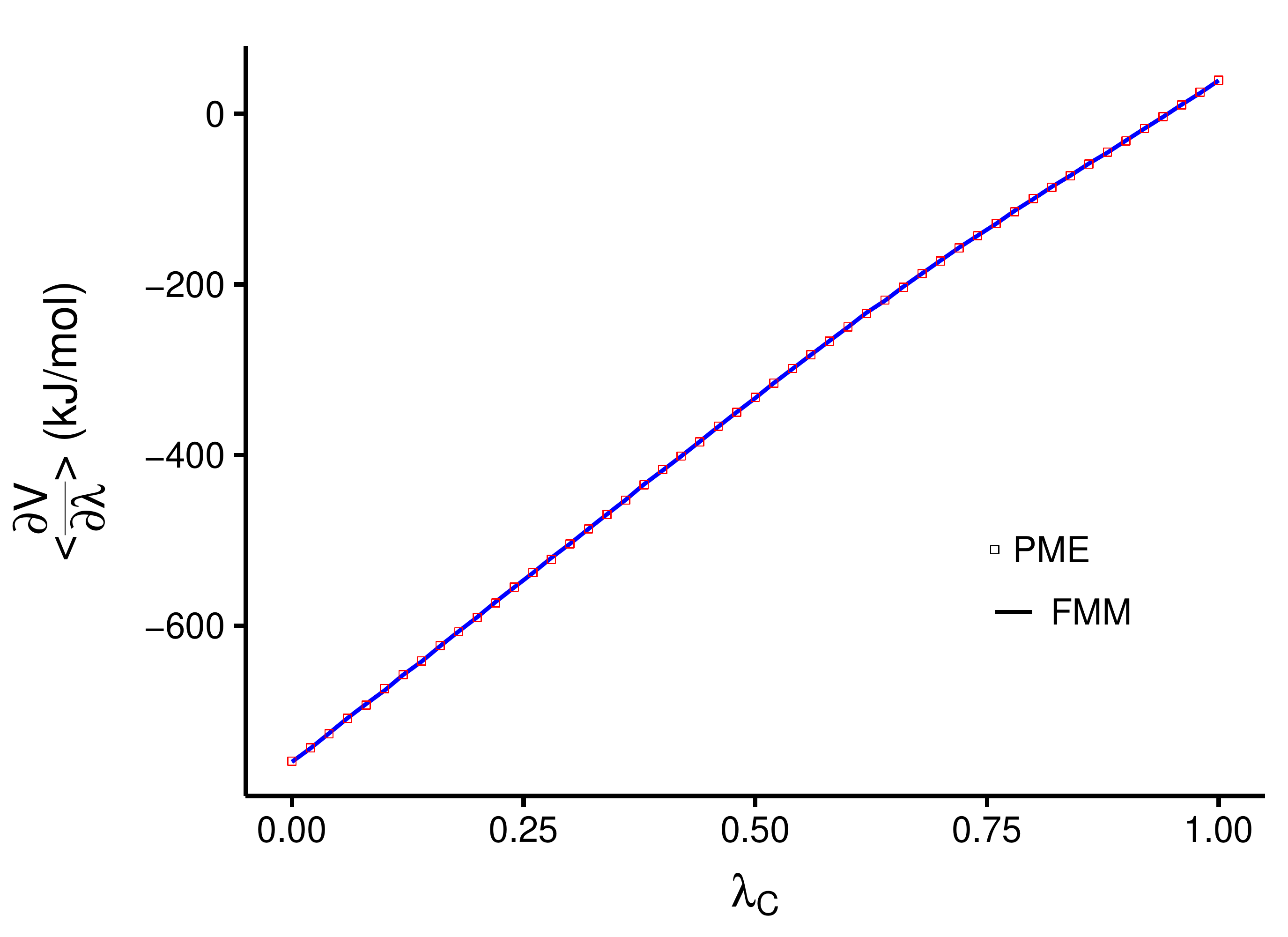}
\caption{\label{Na_average_all}
The calculated \(\frac{\mathrm{d}V}{\mathrm{d}\lambda}\)  with coulombic couplings by FMM (lines) and PME(square) for the removal of Na\(^{+}\) from the solution.}
\end{figure}

\clearpage
\subsection*{\(\frac{\mathrm{d}V}{\mathrm{d}\lambda}\) with LJ coupling analysis for Mg\(^{2+}\), Na\(^{+}\), and Cl\(^{-}\) systems}

The calculated \(\frac{\mathrm{d}V}{\mathrm{d}\lambda }\) with LJ coupling after electrostatic coupling is shown for three ions calculation.

\begin{figure}[htbp]

\includegraphics[width=0.7\textwidth]{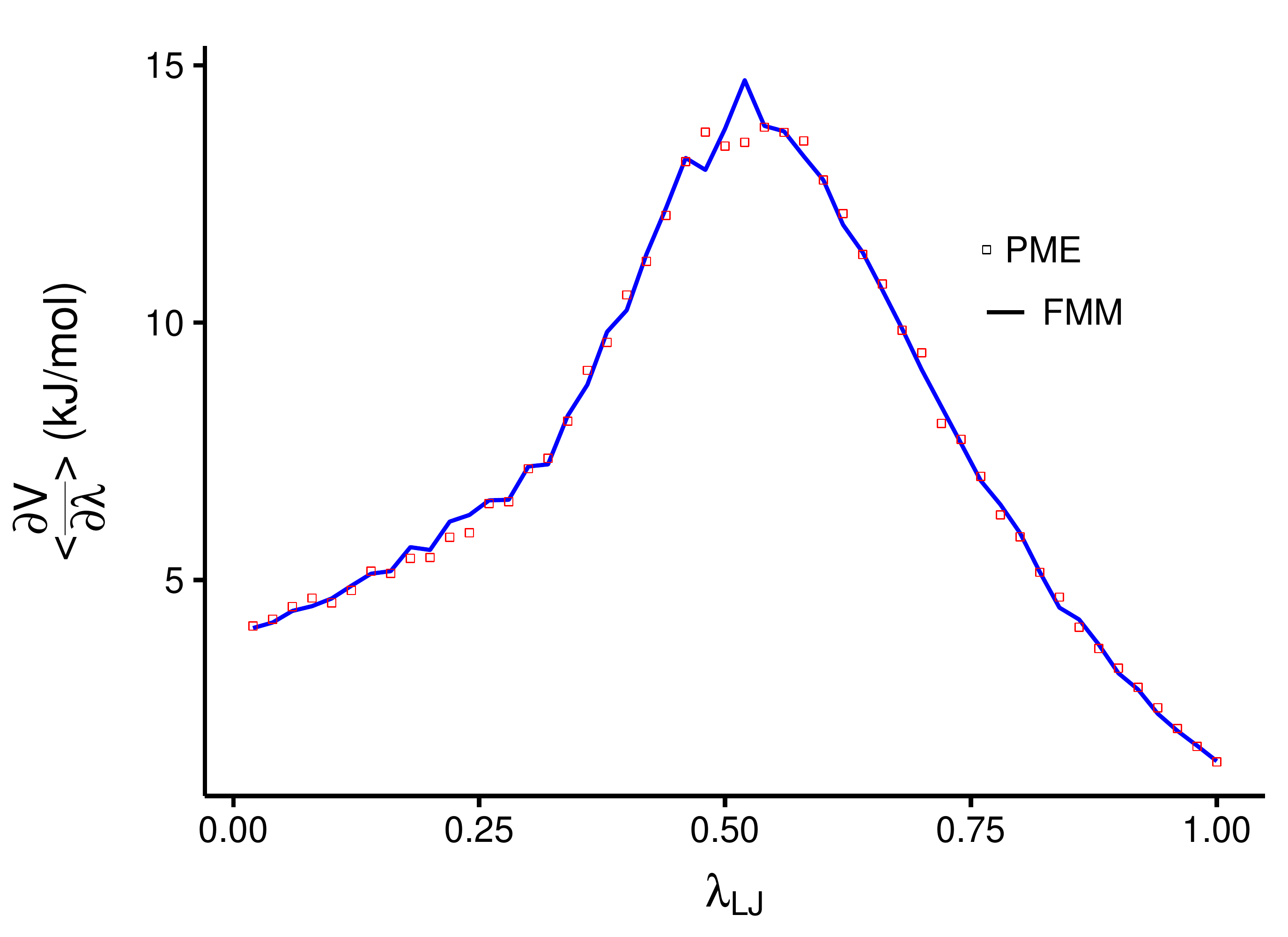}
\caption{\label{Mg_average_lj}
The calculated \(\frac{\mathrm{d}V}{\mathrm{d}\lambda}\)  with LJ couplings by FMM (lines) and PME(square) for the removal of Mg\(^{2+}\) from the solution.}
\end{figure}

\begin{figure}[htbp]

\includegraphics[width=0.7\textwidth]{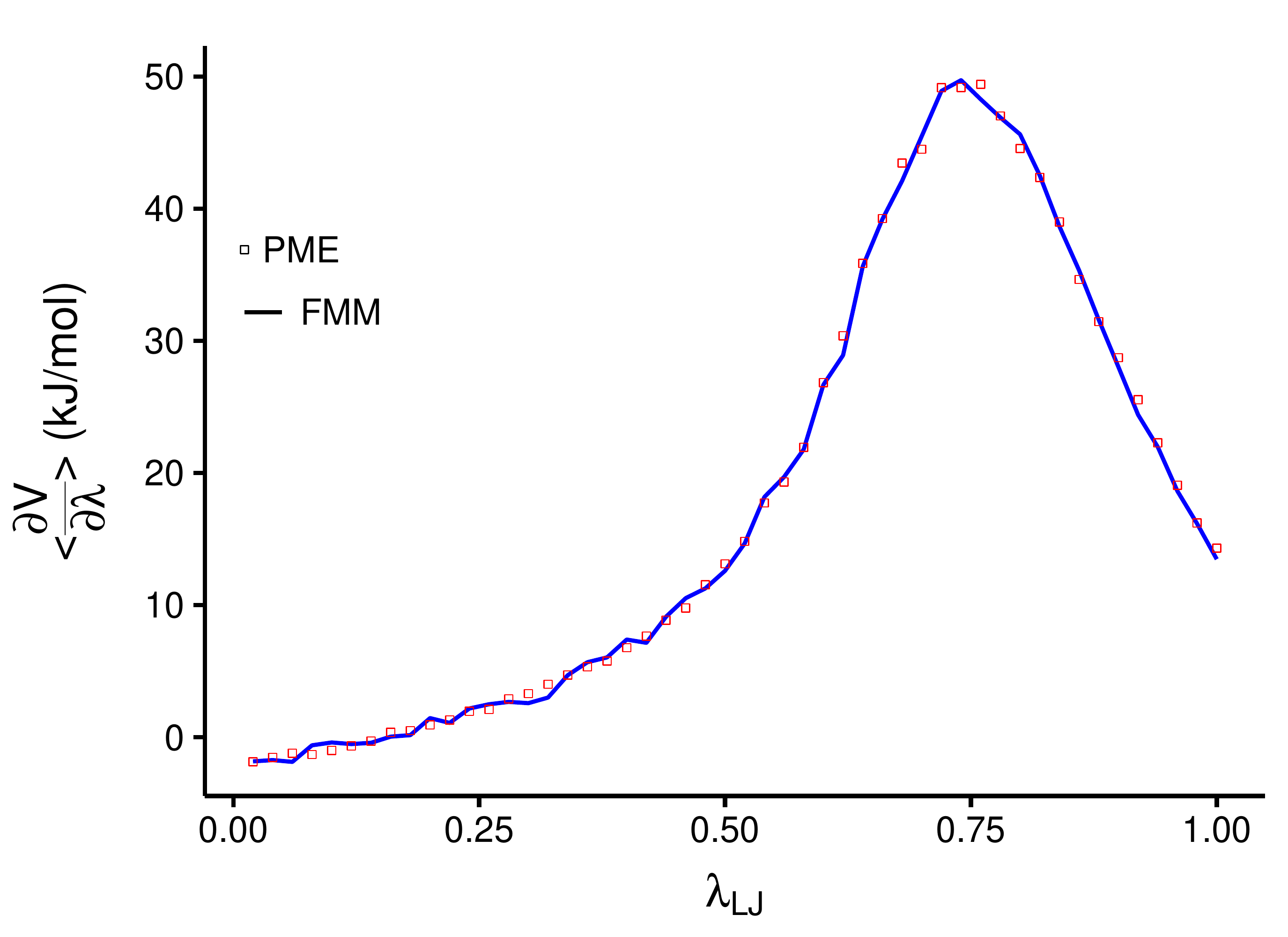}
\caption{\label{Cl_average_lj}
The calculated \(\frac{\mathrm{d}V}{\mathrm{d}\lambda}\)  with LJ couplings by FMM (lines) and PME(square) for the removal of Cl\(^{-}\) from the solution.}
\end{figure}

\begin{figure}[htbp]

\includegraphics[width=0.7\textwidth]{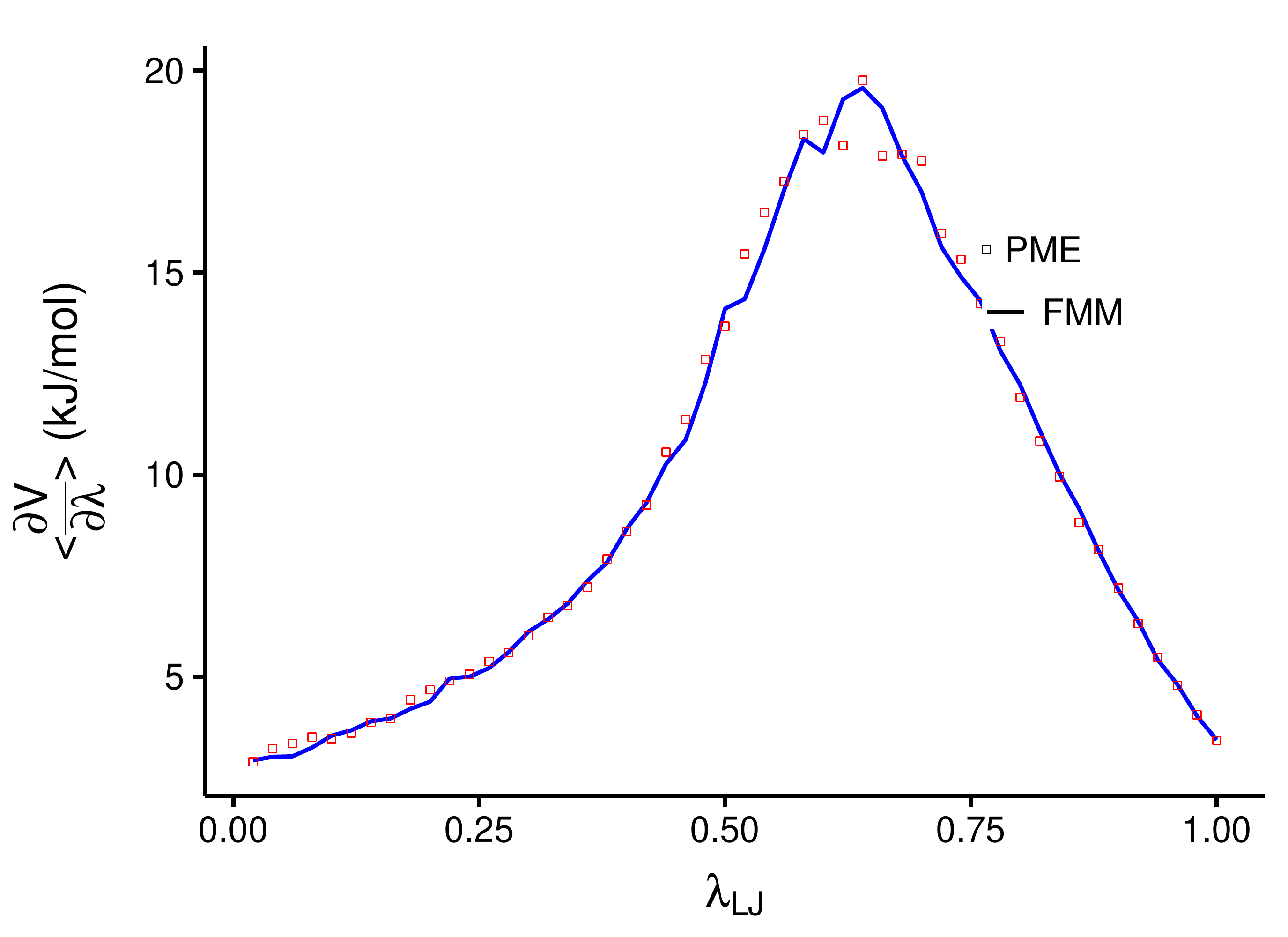}
\caption{\label{Na_average_lj}
The calculated \(\frac{\mathrm{d}V}{\mathrm{d}\lambda}\)  with LJ couplings by FMM (lines) and PME(square) for the removal of Na\(^{+}\) from the solution.}
\end{figure}

\section*{}

\end{document}